\UseRawInputEncoding 
\documentclass[reprint,twocolumn,aps,prb,amsmath,amssymb,floatfix,showpacs]{revtex4-1}

\usepackage[dvipdfmx]{graphicx} 
\usepackage{dcolumn}
\usepackage{bm}
\usepackage{color}
\usepackage[normalem]{ulem} 
\usepackage{graphicx}
\usepackage{bm}
\usepackage{amssymb}
\usepackage[normalem]{ulem} 

\begin{document} 
\title{Nonlinear and Hysteretic Ultrasound Propagation in Solid $^4$He: \\ Dynamics of Dislocation Lines and Pinning Impurities}
\author{Izumi Iwasa$^a$}
\author{Harry Kojima$^b$}
\affiliation{$^{a}$Faculty of Science, Kanagawa University,
Kanagawa, Japan 259-1293
\\$^b$Serin Physics Laboratory, Rutgers University, Piscataway, New Jersey 08854}
\date{\today}

\begin{abstract}
We report on the measurements of 9.6 MHz ultrasound propagation down to 15 mK in polycrystalline quantum solid $^4$He containing 0.3 and 20 ppm $^3$He impurities.  The attenuation and speed of ultrasound are strongly affected by the dislocation vibration.  The observed increase in attenuation from 1.2 K to a peak near 0.3 K is independent of drive amplitude and reflects crossover from overdamped to underdamped oscillation of dislocations pinned at network nodes.  Below 0.3 K, amplitude-dependent and hysteretic variations are observed in both attenuation and speed.  
The attenuation decreases from the peak at 0.3 K to a very small constant value below 70 mK at sufficiently low drive amplitudes of ultrasound, while it remains a high value down to 15mK at the highest drive amplitude.
The behaviors at low drive amplitudes can be well described by the effects of the thermal pinning and unpinning of dislocations by the impurities.  The binding energy between a dislocation line and a $^3$He atom is estimated to be 0.35 K.  The nonlinear and hysteretic behaviors at intermediate drive amplitudes are analyzed in terms of stress-induced unpinning which may occur catastrophically within a network dislocation segment.
The relaxation time for pinning at 15 mK is very short ($< 4$ s), while more than 1,000 s is required for unpinning.
\end{abstract}
\maketitle 

\section{Introduction}
Dislocation lines\cite{Friedel64,Hull65,Nabarro67a} are one dimensional defects in crystal lattice and are important in understanding phenomena such as fracture and fatigue\cite{Suresh1998}, yield strength, plasticity, work hardening, creep, and others.  Though material properties influencing these phenomena may be varied empirically, a complete physical understanding of dislocations is still lacking.  Dislocations in materials can be directly observed by chemical etching, electron transmission microscopy and X-ray diffraction.  The etching is restricted to surfaces; the electron transmission is invasive, local and requires vacuum environment; the X-ray techniques\cite{Tanner76} are limited by low resolution\cite{Iwasa95,Burns08}.
Ultrasound is a sensitive and convenient nondestructive tool for probing the global properties of dislocation lines in materials\cite{Maurel09,Barra15} including solid helium\cite{Franck73,Iwasa79,Beamish82,Lengua90}.  
Ultrasound is especially suited for studying dislocations where their length scale nearly matches the acoustic resonant length.

A network of dislocation lines that are \emph{strongly} pinned at their intersections(nodes) is formed in a solid sample during its growth process.    Dislocation lines may be additionally pinned \emph{weakly} by bound impurity atoms.   According to the Granato-L\"{u}cke theory\cite{Granato56a} (GL) (see Sec. \ref{theory}), under the periodic stress of ultrasound the dislocation lines execute damped vibration like taut strings pinned at their network nodes (average network pinning length $L_{nA}$) and the impurities (average impurity pinning length $L_{iA}$),  and the vibrating dislocation segments modify the ultrasound propagating characteristics. 

Solid $^4$He is particularly suitable for investigating the impurity effects on the dislocation dynamics: the only relevant impurity is the $^3$He atom (nominal concentration is $3\times 10^{-7}$ in commercially available ``natural purity" $^4$He gas).  Though the impurity concentration is very low, the impurities are crucial in understanding the observed phenomena.  At low temperatures($T$) below 1 K, the $^3$He impurity atoms are highly mobile owing to the rapid quantum diffusion process\cite{Schratter84} and uniformly distributed within a solid $^4$He sample.   $^3$He impurity atoms lower their elastic energy by ``condensing" onto (predominantly edge) dislocations and thereby pin dislocation lines.  The dislocation lines, however, can be unpinned from the impurity atoms by increasing the amplitude of the ultrasound stress when the imposed force on the pinning site exceeds a critical force.  
Once unpinning is initiated on a dislocation line, it continues at all other impurity atoms on the line till this line is only pinned at its two end network nodes. This runaway effect was anticipated by GL\cite{Granato56a,Kang15} as ``catastrophic break-away" of dislocation lines from \emph{immobile} impurities.  
An important difference in our case is that the impurities are \emph{mobile}.  When unpinning occurs in solid $^4$He, the mobile impurities rapidly diffuse away from the dislocation and cannot repin the dislocation again within the acoustic cycle.

Franck and Hewko\cite{Franck73} reported the first measurement of temperature dependence of the ultrasound speed in single crystals of hcp $^4$He above 0.7 K.  They discovered a transition in the speed from a power law expected for the adiabatic sound at high temperatures to approximate temperature independence below 1.1 K or 2 K depending on the growth pressure of crystals.  Wanner, et al.\cite{Wanner76} made similar measurements down to 0.2 K and found that the deviation from the adiabatic sound speed at temperatures below 1 K could be both positive and negative depending on samples.  They identified the origin of the observed anomalous temperature dependence below 1 K as the resonant interaction of dislocation lines with ultrasound. Since then, the effects of dislocations on the propagation speed as well as attenuation of ultrasound in single-crystal solid helium have been confirmed by other studies\cite{Iwasa79,Iwasa80,Beamish82} down to 80 mK.  The effects of pinning by dislocations in solid $^4$He doped with additional $^3$He have been studied\cite{Iwasa80} in the higher temperature range.

The interaction between the dislocation lines and $^3$He impurities in hcp $^4$He was also revealed in the measurements of shear modulus in the kHz range\cite{Day07,Haziot13b,Fefferman14,Kang13}.  These measurements were originally stimulated by the observation of the so-called NCRI (non-classical rotational inertia) in the torsional oscillator experiments\cite{Kim04a,Kim04b}.
The variations of the shear modulus and dissipation at tempratures below 1 K are found to depend on the frequency, strain amplitude and temperature.  The mechanism of the interaction is either damping of dislocation motion or pinning of dislocation lines by $^3$He impurities.  Parameters related to the dislocation such as the dislocation density, the average network pinning length, the damping constant of dislocation motion, and the binding energy of a $^3$He atom to the dislocation are obtained from these measurements.

The present work is the first to study the ultrasound propagation in \emph{polycrystalline} solid $^4$He extended down to 15 mK.  Typically, the attenuation and speed of ultrasound increase as the sample is cooled from 1.2 to 0.3 K.  These responses are similar to those in the single-crystalline samples and can be explained by the transition from overdamped to underdamped resonance of dislocations.  
The ultrasound responses become nonlinear and hysteretic as the sample is further cooled.  When the drive level is sufficiently low, the attenuation typically decreases by 20 dB from 300 to 15 mK, while the speed shows a minimum at 100 mK.  These low drive responses will be explained as an effect due to pinning of the dislocations by $^3$He impurities in thermal equilibrium.  When the drive is sufficiently high, on the other hand, the attenuation decreases very little upon cooling from 300 to 15 mK.
Hysteretic responses are observed depending on the history of how the drive level and temperature are varied.  For example, the attenuation in the warming run at an intermediate drive level is smaller than the corresponding attenuation in the cooling run at the same drive level.  The increase in attenuation in the warming run will be shown to originate in the stress-induced ``catastrophic" unpinning of dislocation lines from $^3$He impurities.  These observations of nonlinear and hysteretic responses are made for the first time in the ultrasound measurement on solid $^4$He with 0.3 ppm of $^3$He.

The paper is organized as follows.  The experimental methods are described in Sec. \ref{apparatus}.  The data on the measured attenuation and speed on typical as well as atypical samples are shown in Sec. \ref{Results}.
In Sec. \ref{Analysis} the theoretical background on the GL theory is first given.  Then the behavior of the typical sample is analyzed.  Discussions on the atypical samples and comparison with other related experiments are given in Sec. \ref{Discussion}.  The paper is concluded with a brief summary and questions for further research in Sec. \ref{Conclusion}.  Supplemental information is given in the Appendices.

The analysis is rather complicated because of the nonlinear nature of the phenomena.  There are two independent external parameters, i. e. temperature and stress, which affect pinning and unpinning of dislocations by $^3$He impurities.  The temperature of the sample can be regarded as uniform all over the sample.  The stress, on the other hand, is inhomogeneous because the amplitude of the ultrasound pulse decreases along the path of ultrasound due to attenuation.  Therefore, a position dependent analysis carried out is described in detail in the latter part of Sec. \ref{Analysis}.

\section{Apparatus and Measurement Procedure}\label{apparatus}
\begin{figure}
\includegraphics[width=3.4in]{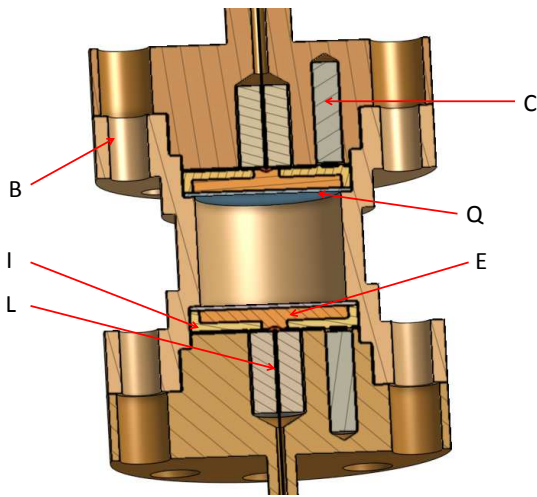}
\caption{Cutaway schematic view of ultrasound cell: quartz transducer(Q), electrical contact plate to transducer(E), insulating backing (I), one compression spring(C) of three, electrical lead(L) attached to spring-loaded contact, bolt holes(B, bolts not shown).  The brass cell body encloses the cylindrical sample chamber between the transducers.  Helium is filled via the hole for the upper electrical lead.  The cell is attached to the mixing chamber of the dilution refrigerator via the upper BeCu (torsion) rod with the sample chamber axis oriented vertically.}
\label{schematic}
\end{figure}

Our ultrasound apparatus (a schematic is shown in Fig. \ref{schematic}) for studying solid $^4$He has been described earlier.\cite{Hein13}  It was originally designed to study the ultrasound propagation simultaneously with torsional oscillation response of solid $^4$He.  This report is focussed on the ultrasound results.  The interrelationship between the ultrasound and the torsional oscillation phenomena will be described elsewhere.  Briefly, a BeCu torsion rod (outer diameter = 4.0 mm, inner diameter = 0.8 mm, length = 14 mm) connected to the cylindrical sample chamber provides both the thermal contact to the mixing chamber of a dilution refrigerator and an inlet conduit for the sample gas.  The inner ends of the sample chamber (diameter($D$) = 8.6 mm, length($x_m$) = 6.6 mm) are terminated by identical (10 mm diameter, 10 MHz) X-cut quartz transducers acting as driver and detector.  The sample length is set by the precision-machined steps on the inner wall of the sample chamber as shown in Fig. \ref{schematic}.  Ultrasound is excited by applying 1.2 $\mu$s wide RF voltage pulses of amplitude $V$ and frequency $\Omega/2\pi = 9.6$ MHz on the driver transduder with a repetition frequency of 1 kHz, where $\Omega$ is the angular frequency.  The ultrasound relative drive amplitude $A \equiv V/V_0$ is normalized to an arbitrary value of $V_0$ = 2.53 mV.
As described in Appendix \ref{AppendixA} on calibration of transduers, $A$ = 1 is estimated to be equivalent to an applied compressional stress amplitude of 1.36 Pa.  The estimation is, however, uncertain by as much as a factor of two due to incomplete impedance matching, acoustic mismatch, and ringing of the transducer in the ultrasound experiment.

A standard pulse-echo apparatus with a superheterodyne system\cite{Lengua90} is used to measure the ultrasound propagation.  The received pulse signal is converted to an intermediate frequency of 5 MHz, amplified, and phase-sensitively detected in the spectrometer. The resulting quadrature video voltages, $V_{\sin}$ and $V_{\cos}$, are then measured by a boxcar integrator.  Fractional changes in propagation speed $v$ are evaluated from the phase shift according to: $\delta v/v =\arctan (V_{\sin}/V_{\cos})/(\Omega P)$ where $P = x_m/v$ is the transit time.  The relative attenuation of sound signal amplitude (in dB) is evaluated according to $\alpha = -20\log\sqrt{V_{\sin}^2 + V_{\cos}^2} - \alpha_v$, where  $\alpha_v$ is the drive output attenuator setting (in dB).   The zero offsets of the quadrature signals are determined in the presence of solid $^4$He sample in the cell by extrapolating the measured voltages as the drive amplitude is decreased nearly to zero.
The absolute speed of sound can be in principle estimated from the length of the sample and the transit time of the ultrasound pulse in the pulse-echo method.  Unfortunately, we could not accurately measure the absolue speed because the rising edge of the received signal was rounded due to ringing of the transducer.

\begin{figure*}
\begin{center}
\includegraphics[width=7.0in]{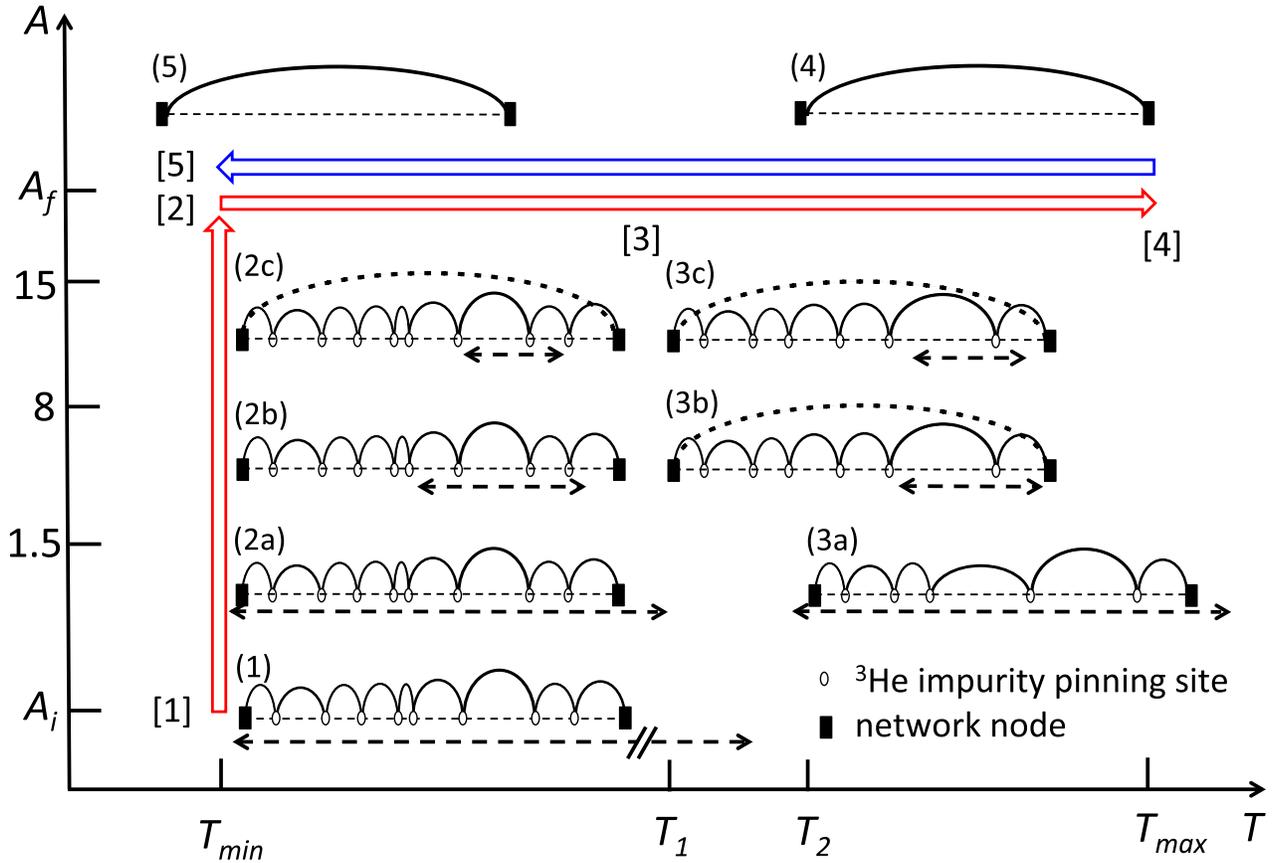}
\caption{
Measurement procedure of varying the ultrasound relative drive amplitude $A$ and temperature $T$ for the sample $\#11$.  The axis tick marks are not to scale.  At $T_{\min}$ ($\sim$15 mK) the drive amplitude is increased from $A_i \leq 0.07$ at [1] to $A_f = 1 - 15$ at [2].   The temperature is subsequently increased in a warming run in $\sim$6 hours from [2] to $T_{\max}$($\sim$1.2 K) at [4].   
The temperature is then decreased over $\sim$15 hours from [4] to [5] in a cooling run.  The procedure is completed by decreasing $A$ down to $A_i$ ([5]$\rightarrow$[1]).  The next run is repeated for a different $A_f$.  Characteristic temperatures $T_1$ and $T_2$ are described in the text(Sec. \ref{T_A_dep}).  Cartoons (1)$-$(5) illustrate pinning/unpinning and string-like deflections of one network segment of dislocation pinned by impurities.  Heavy dashed horizontal lines with arrows indicate the critical length $L_c$ as $A_f$ is varied.  Heavy dashed curves connecting the two end network nodes indicate that the network segment is catastrophically unpinned(see Sec. \ref{catastrophic}) from impurities.
}
\label{procedure}
\end{center}
\end{figure*}

Solid $^4$He samples are grown in the cylindrical chamber by the blocked capillary method as follows.  The chamber is initially loaded at 4.2 K with liquid $^4$He at 68 bar, using commercial natural purity $^4$He gas originated from Texas with a nominal $^3$He impurity concentration($x_3$) of 0.3 ppm\cite{Souris15}.  As cooling of the dilution refrigerator system is initiated, the section of the fill capillary attached to the high temperature end of the refrigerator rapidly cools down near to 1.2 K.  The liquid $^4$He within that section of the capillary becomes frozen and forms a plug, which now keeps the $^4$He mass below the plug and in the sample chamber fixed.  Subsequently, an ``as-grown" solid sample with a molar volume of 20.3 cm$^3$/mole under a final pressure of about 35 bar is produced over about two hours during which the sample chamber is cooled below 50 mK.  

Annealing is carried out in some samples by raising and maintaining a constant temperature  near 1.5 K for 20 - 48 hours.  The sample pressure is indirectly monitored by the pressure sensor located on the mixing chamber.  The sample chamber and the pressure sensor is connected by the fill line tube (inner diameter = 0.75 mm and length 15 cm).  The sensor pressure is found to decrease monotonically during annealing.  Annealing is considered completed when the changes in the pressure sensor reading are much reduced from the initial rate.

The effects of $^3$He impurity concentration on ultrasound response are studied by increasing $x_3$ to 20 ppm.  The dilution refrigerator system is warmed up to about 5 K and the remnant commercial natural purity helium gas from the previous measurements is pumped out from the sample cell over 22 hours.  The sample cell is next loaded with the appropriate amount of $^3$He gas to make up the increased $x_3$, filled and finally pressurized with commercial natural purity $^4$He at 4.2 K.  The mixture solid sample is subsequently grown in the usual manner described above.  The \emph{in situ} $^3$He concentration within the sample solid itself is not measured.

Measurements on the samples grown at the initial stages of the experiment were made as a function of temperature at relatively high drive amplitudes, $A >$ 2.5.  The measurements showed that the ultrasound response became nonlinear at low temperatures, i. e., both $\alpha$ and  $\delta v/v$ were amplitude dependent and hysteretic at temperatures below 0.3 K.  It was finally found that decreasing $A$ below a critical value was crucial in allowing $^3$He impurities to pin the dislocations in the sample and suppressing the nonlinear effects.  In order to study the details of the effects related to changing $A$, the measurement procedure illustrated in Fig. \ref{procedure} was adopted for the last and most extensively studied sample $\#$11.

The measurement procedure begins at a minimum temperature $T_{\min}\sim 15$ mK where the relative drive amplitude is set at an initial low amplitude $A_i (\leq 0.07)$ (see [1] in Fig. \ref{procedure}).   This $A_i$ is sufficiently small that all the dislocation segments are pinned by the number of $^3$He impurities in thermal equilibrium.  Keeping the temperature constant, $A$ is then increased in several steps to the final relative amplitude $A_f$ at [2] in Fig. \ref{procedure}.  The temperature is then increased in a ``warming run" in $\sim6$ hours ([2]$\rightarrow$[3]$\rightarrow$[4]) to $T_{\max}\sim 1.2$ K, where the temperature is held constant for about 20 minutes.  The temperature is then decreased in a ``cooling run" back down to $T_{\min}$ over $\sim15$ hours ([4]$\rightarrow$[5]).   At $T_{\min}$ $A$ is decreased down to $A_i$([5]$\rightarrow$[1]) and then increased to a different $A_f$ for the next run.  The ultrasound data, $\delta v/v$ and $\alpha$, are acquired throughout the procedure.  Depending on $A_f$, the ultrasound data can be hysteretic or reversible during the warming and successive cooling run as described below.

\section{Results}\label{Results}
Ultrasound propagation was studied in eleven solid $^4$He samples in hexagonal close-packed (hcp) structure.  Nine of them were grown from commercial $^4$He gas ($x_3$ = 0.3 ppm).  Two samples with $x_3$ = 20 ppm were grown in order to examine the effect of $^3$He impurity on ultrasound response.  The  ultrasound response varied in detail from sample to sample.  Characteristics of the samples, however, could be roughly divided into three categories depending on the nature of observed ultrasound response. The samples described in this paper are listed in Table \ref{fit_parameters2} together with fitted parameters to be discussed later.
 
Many of the samples show ``typical" ultrasound response that follows from the interaction of the ultrasound with dislocation lines and is compatible with the GL theory\cite{Granato56a} when pinning at the network nodes and by $^3$He impurities is taken into account.  The typical samples have an attenuation peak (typically 20 dB) around 0.3 K.  The attenuation and the change in ultrasound speed become nonlinear and hysteretic below the peak temparature.  Samples with ``small attenuation'' belong to the second category.  Their attenuation peak is less than 3 dB and the variation of the speed of ultrasound is almost proportional to $T^4$, that is, the response is almost free of dislocation effects.
In the third rare category, samples exhibit ``anomalous" response where the ultrasound response around 0.7 K is of a relaxation type.  This response cannot be simply explained by the GL theory.  However, the nonlinear and hysteretic effects of dislocations similar to the typical samples can be seen also in the anomalous samples at temperatures below 100 mK.
Taken together, our samples allow a comprehensive study of solid $^4$He exhibiting a wide range of ultrasound propagation phenomena produced by dislocations.

\subsection{Typical sample response}

\subsubsection{Temperature and drive amplitude dependence}\label{T_A_dep}

\begin{figure}
\includegraphics[width=3.4in]{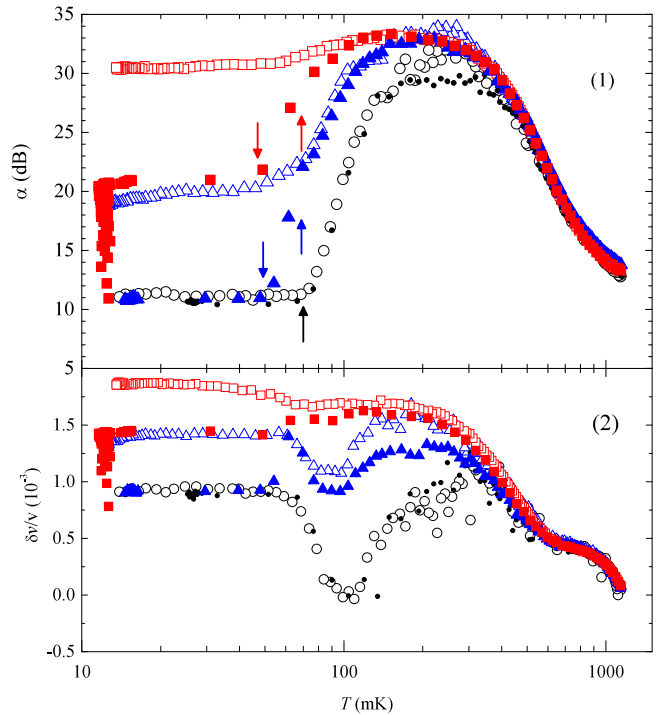}
\caption{Temperature dependence of $\alpha$ (panel (1)) and $\delta v/v$ (panel (2)) of the sample $\#11$ with 0.3 ppm of $^3$He impurity during warming(w) and subsequent cooling(c) runs with three different drive amplitudes.  $A_f$ =  1.12 (black closed(w) and open(c) circles), 3.98 (blue closed(w) and open(c) triangles), and 14.1 (red closed(w) and open(c) squares).   Down(up) arrows in (1) indicate characteristic temperatures $T_1$($T_2$) where the dislocation line pinning state changes in warming runs.  See text(Sec. \ref{T_A_dep}) for explanation of these characteristic temperatures and the  ``vertical" data at $T_{\min}$ for $A_f = 14.1$.}
\label{L1-2D-T-3drive-atten-dv-wc-v2}
\end{figure}

Temperature dependences of $\alpha$ and $\delta v/v$ of a typical sample taken during warming and subsequent cooling runs with three values of drive amplitude ($A_f$=1.12, 3.98, and 14.1) are shown in Fig. \ref{L1-2D-T-3drive-atten-dv-wc-v2}.   The data were taken on the sample $\#11$ (melting temperature 1.9 K) after annealing at 1.5 K for 22 hours. Before starting each warming run on this sample, the drive amplitude was decreased to  $A_i \leq 0.07$ and then increased to a new $A_f$ according to the measurement procedure in Fig. \ref{procedure}.
When the drive amplitude was decreased to $A_i$ at $T_{\min}$, $\alpha$ and $\delta v/v$ became $\alpha _0=10$ dB and $c=0.9\times 10^{-3}$, respectively, regardless of the previous value of $A_f$.  We regard $\alpha_0$ and $c$ as the reference values for $\alpha$ and $\delta v/v$.

There is a broad attenuation peak around 200 $\sim$ 300 mK. In the high temperature range, $T > 300$ mK, the response is linear: both $\alpha$ and $\delta v/v$ are independent of $A_f$ as well as the temperature sweep direction. The values of $\alpha$ and $\delta v/v$ decrease monotonically as the temperature increases.  It will be shown below (see Sec. \ref{LinearRegime}) that the interaction between the ultrasound and dislocation lines pinned at network nodes can account for the behavior in this temperature range.

In the low temperature range, $T < 200$ mK, the response becomes more intricate. It is strongly nonlinear, i.e. dependent on the value of $A_f$, and hysteretic with respect to the thermal history.  The attenuation increases monotonically from $T_{\min}$ to 200 mK, whereas $\delta v/v$ shows a distinct minimum around 100 mK and occasionally a maximum at a lower temperature. 
On cooling runs, both $\alpha$ and $\delta v/v$ tend to saturate to $A_f$-dependent values as $T_{\min}$ is approached.  The low-temperature response will be analyzed in terms of pinning of dislocations by $^3$He impurities in Sec. \ref{ImpurityPinning} and below.

On warming at $A_f$=1.12, $\alpha$ remains equal to $\alpha _0$ up to a characteristic temperature $T_2(\approx 70$ mK), above which it increases to reach the broad maximum.  On cooling at $A_f$=1.12, $\alpha$ decreases from the broad peak and reaches $\alpha _0$ at $T_2$.  There is no hysteresis between the warming and cooling runs.  Similar reversible behavior was observed in another run at $A_f$=1.41 (not shown in Fig. \ref{L1-2D-T-3drive-atten-dv-wc-v2}).
The variation of $\delta v/v$ at $A_f$=1.12 is also thermally reversible and the minimum around 100 mK is the deepest among the runs in Fig. \ref{L1-2D-T-3drive-atten-dv-wc-v2} (2).

The attenuation at at $A_f$=3.98, on the other hand, is hysteretic.  On warming, $\alpha$ remains equal to $\alpha _0$ up to another characteristic temperature $T_1(\approx 50$ mK), above which it increases towards the broad maximum with a small kink at $T_2$.  On cooling, $\alpha$ decreases from the maximum along the warming data down to $T_2$, where it departs from the warming line and levels off.
The variation of $\delta v/v$ at $A_f$=3.98 is also hysteretic.  Note that there is a maximum in $\delta v/v$ on the warming run around $T_2$.

At $A_f$=14.1, $\alpha$ already increases at $T_{\min}$ above $\alpha_0$ as shown by ``vertical points" in Fig. \ref{L1-2D-T-3drive-atten-dv-wc-v2}(1) and reaches 21 dB over about 2000 s time interval after $A$ is changed from $A_i$ to $A_f$.  The temporal variations will be described in Sec. \ref{relaxation}.
When the temperature is raised, $\alpha$ remains constant at 21 dB up to $T_1$ and then increases towards the broad maximum.  On cooling at $A_f$=14.1, $\alpha$ decreases only slightly from the maximum value.
The behavior of $\delta v/v$ at $A_f$=14.1 shows similar features; $\delta v/v$ increases slowly at $T_{\min}$ after $A$ is changed from $A_i$ to $A_f$, and the variations in $\delta v/v$ between 50 and 200 mK are smaller than those at lower $A_f$.

\begin{figure}
\includegraphics[width=3.4in]{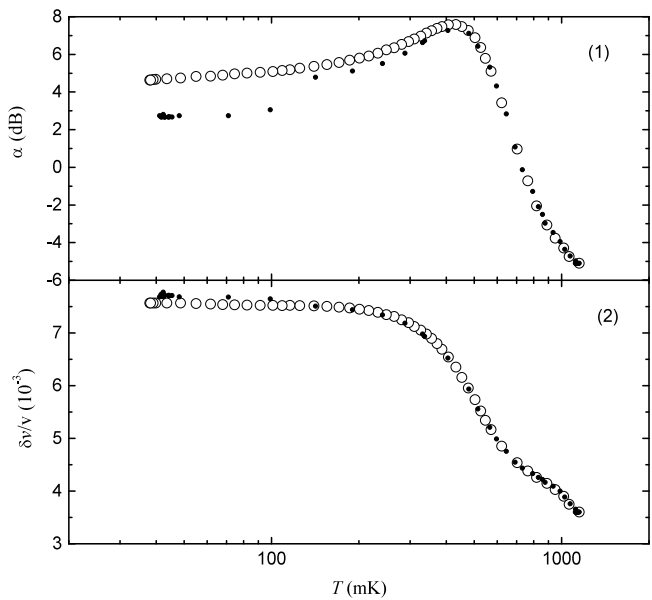}
\caption{Attenuation (panel (1)) and change in speed (panel (2)) of the sample $\#1$ with 0.3 ppm 
of $^3$He impurity in warming (black dots) and cooling (open circles). The drive level is high ($A_f$=7.9) and the measurement procedure deviates slightly from Fig. \ref{procedure} (see text).}
\label{sampleA1-23db}
\end{figure}

Figure \ref{sampleA1-23db} shows $\alpha$ and $\delta v/v$ of another typical sample $\#1$, the first sample in this experiment. The sample was produced by annealing an as-grown sample at 1.53 K for 24 h. The procedure of measurements was different from that in Fig. \ref{procedure}. Before starting the warming run, the drive amplitude was increased from $A_f$ = 2.51 of the previous run to $A_f$ = 7.94 without reducing to $A_i$.  As a result, $\alpha$ and $\delta v/v$ in the warming run did not start from $\alpha _0$ and $c$, respectively.
The variations of $\alpha$ and $\delta v/v$ in the cooling run are similar to those of the sample $\#11$ at $A_f$ = 14.1 shown in Fig. \ref{L1-2D-T-3drive-atten-dv-wc-v2}.

\subsubsection{Hysteresis with respect to drive amplitude}\label{hysteresis_section}
\begin{figure}
\begin{center}
\includegraphics[width=3.4in]{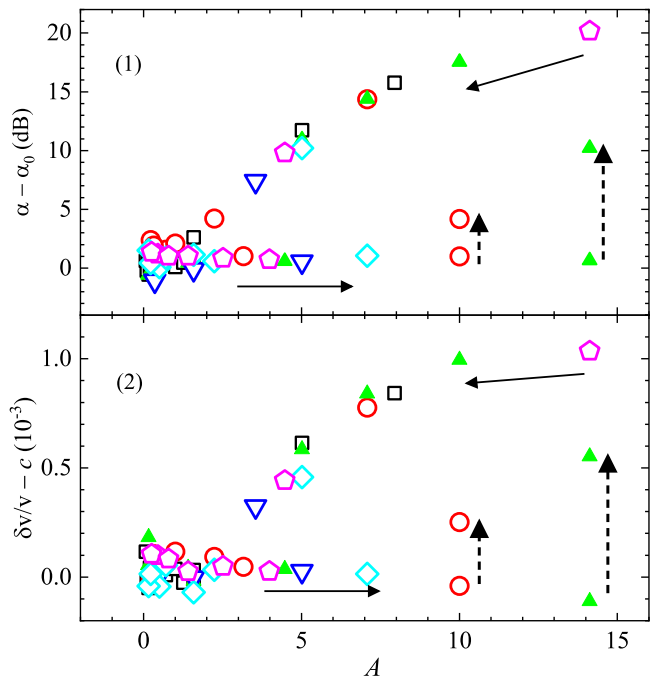}
\caption{Hysteresis of attenuation $\alpha - \alpha_0$ (panel (1)) and speed $\delta v/v - c$ (panel (2)) vs. drive amplitude $A$ at $T_{\min}$ in the same sample $\#11$ as in Fig. \ref{L1-2D-T-3drive-atten-dv-wc-v2}.  
The drive amplitude is decreased (solid arrows pointing left) from $A_f$ at the end of previous cooling run (Fig. \ref{procedure}[5]) to $A_i$ (Fig. \ref{procedure}[1]) and then increased (solid arrows pointing right) to $A_f$ of the next warming run (Fig. \ref{procedure}[2]).
Different symbols indicate various sequential changes in $A$ starting from different $A_f$ at the end of previous cooling run.  The vertical dashed arrows indicate time dependent response (see Sec. \ref{relaxation}). }
\label{hysteresis}
\end{center}
\end{figure}

According to the procedure prescription shown in Fig. \ref{procedure}, $A$ is decreased at the end of the previous cooling run at $T_{\min}$ from $A_f$ to $A_i$ and then increased to a new $A_f$ for the next warming run.
Figure \ref{hysteresis} shows the ultrasound response, $\alpha - \alpha_0$ and $\delta v/v - c$, of the sample $\#11$ during this process.  Data after subtracting the reference values $\alpha_0$ and $c$ are plotted for clarity here.  Each symbol represents a series of changes taken in $A$ at the end of a cooling run.  It can be seen that different symbols trace out the same hysteretic dependence on $A$ regardless of the value of $A_f$ of the previous cooling run.  

In the process of decreasing $A$ down to a critical value $A_c(\approx 1.5)$, the response is strongly dependent on $A$ indicating nonlinear behavior.   In the range $A_i < A < A_c$, both $\alpha - \alpha_0$ and $\delta v/v - c$ become zero and the response is linear.  In the process of increasing $A$ but only up to a threshold drive amplitude $A_t (\approx 8)$, $\alpha - \alpha_0$ and $\delta v/v - c$ remain zero and the response is linear.  When $A$ is increased to greater than $A_t$, the response becomes nonlinear and time dependent behavior is observed (see the next subsection).

\subsubsection{Time dependent response after drive amplitude change}\label{relaxation}
In \textit{all} of the decreasing steps of $A$ and in the increasing steps of $A$ up to $A_t$ in Fig. \ref{hysteresis}, the changes in response occur rapidly.  However, when $A$ is increased to that above $A_t$, changes in both $\alpha$ and $\delta v/v$ occur in a remarkable transient manner accompanied by long relaxation times indicated by the vertical dashed lines in Fig. \ref{hysteresis}. 

\begin{figure}
\includegraphics[width=3.4in]{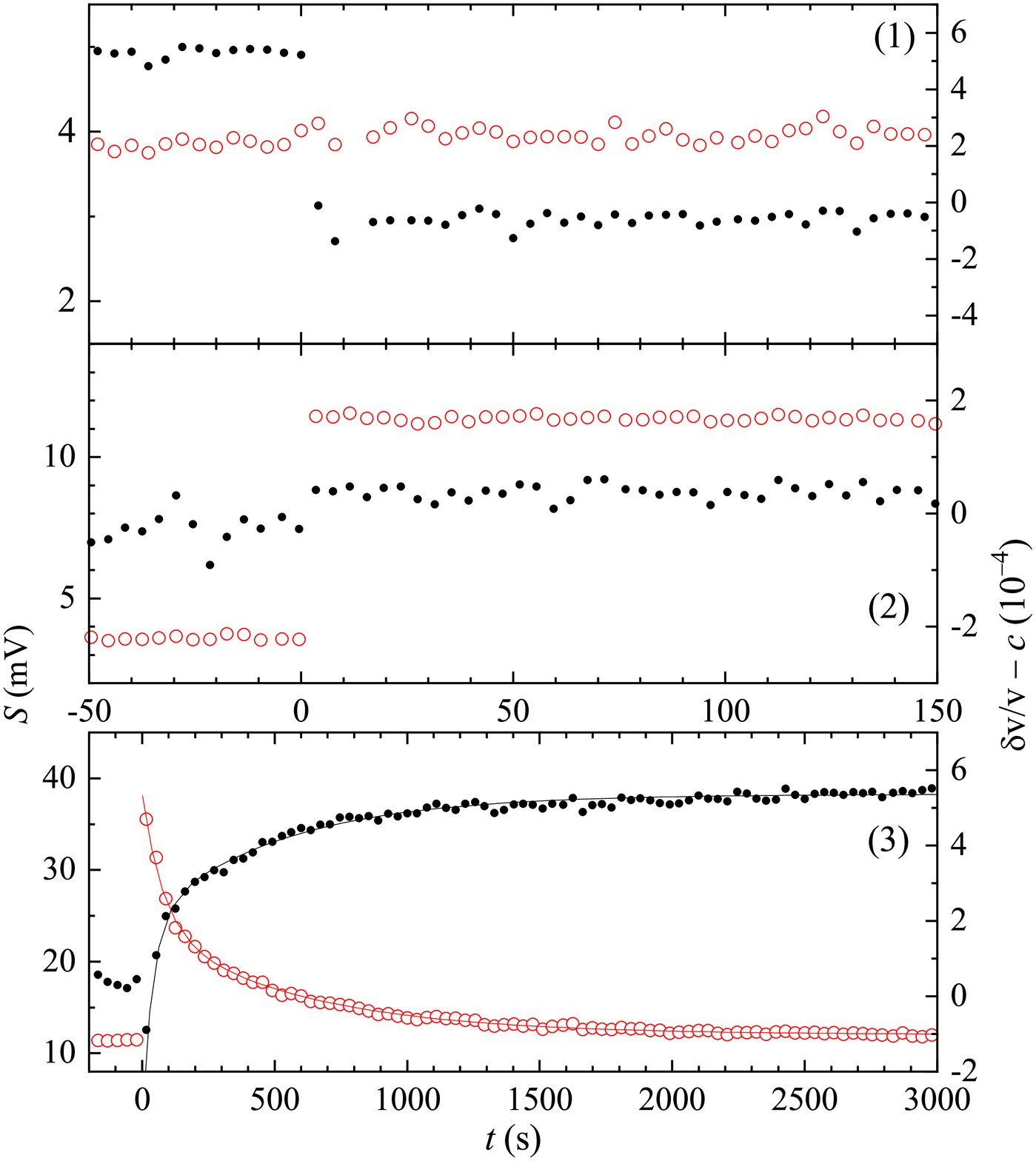}
\caption{
Temporal variations of ultrasound amplitude S (left ordinate, red open circles) and speed $\delta v/v - c$ (right ordinate, black dots) before and after step changes in drive amplitude $A$ (at $t = 0$) in the sample $\#11$ at  $T_{\min}$. Panel (1): $A = 5.01 \rightarrow 1.58$, panel (2): $A =1.41 \rightarrow 4.47$ and panel (3): $4.47 \rightarrow 14.1$.  The number of plotted data points is reduced in (3) for clarity.  The observed relaxations of $S$ and $\delta v/v - c$ in (3) can be fitted with double exponential decays as shown by the curves (see Sec. \ref{relaxation}).}
\label{amp-dv-relaxation}
\end{figure}

Figure \ref{amp-dv-relaxation} shows examples of the ultrasound response at $T_{\min}$ as a function of time($t$) after $A$ is changed at $t=0$.  The amplitude of the received signal, $S = \sqrt{V_{\sin}^2 + V_{\cos}^2}$, and $\delta v/v - c$ are plotted.  In Fig. \ref{amp-dv-relaxation}(1), $A$ is decreased by 10 dB from 5.01 to 1.58 at $t=0$.  No change in $S$ occurs at t=0, indicating that the response is nonlinear and that the attenuation decreases by 10 dB.  The speed of sound, on the other hand, decreases by $6 \times 10^{-4}$ at $t=0$.
The response occurs rapidly within the data acquisition cycle time of $\sim$4 s.

When $A$ is increased by 10 dB from 1.41 to 4.47 (Fig. \ref{amp-dv-relaxation}(2)), $S$ instantly increases by 9.8 dB (3.7 to 11.5 mV) while  the change in $\delta v/v$ is less than $1 \times 10^{-4}$.  These responses are therefore nearly linear.

When the drive is stepped up to one above $A_t$, however, the response is dramatically different.  The drive is increased by 10 dB from 4.47 to 14.1 at $t=0$ in Fig. \ref{amp-dv-relaxation}(3).  $S$ initially increases by 10 dB from 11.5 to 35.7 mV but then decreases slowly, whereas  $\delta v/v - c$ initially becomes negative and then gradually increases to become positive. 
These responses correspond to the vertical dashed lines at $A = 14.1$ in Fig. \ref{hysteresis} (1) and (2) and ``vertical" data at $T_{\min}$ in Fig. \ref{L1-2D-T-3drive-atten-dv-wc-v2}.  The decrease in $S$ cannot be described by a single exponential function.  The fitted curve to $S$ in Fig. \ref{amp-dv-relaxation}(3) is a double exponential function of the form, $S = s_1 + s_2\exp(-t/\tau_1) + s_3\exp(-t/\tau_2)$, where $s_1 = 11.8$ mV, $s_2 = 15.5$ mV, $s_3 = 11.0$ mV, $\tau_1 = 90$ s, and $\tau_2 = 640$ s.  The value of $S$ after $t=2000$ s is almost the same as that before $t=0$, that is, the attenuation increases by 10 dB. The relaxation in $\delta v/v$ can be similarly fitted by a double exponential function.

The observed time dependence of the ultrasound response following step changes in $A$ is most likely revealing the characteristic time scale in the dynamics of the interaction between ultrasound and dislocation lines.  The time scale rapidly decreases from 2000 s at $T_{\min}$ to below 10 s at 65 mK. Unfortunately, detailed temperature dependence of the relaxation phenomenon was not measured.

We will discuss later (see Sec. \ref{relaxation_phenomena}) the physical mechanism behind the nonlinear effects observed in Fig.\ref{amp-dv-relaxation} (1) and (3) in terms of the dynamic process of pinning and unpinning of dislocation lines by $^3$He impurities. 

\subsection{Small-attenuation samples}
\begin{figure}
\includegraphics[width=3.4in]{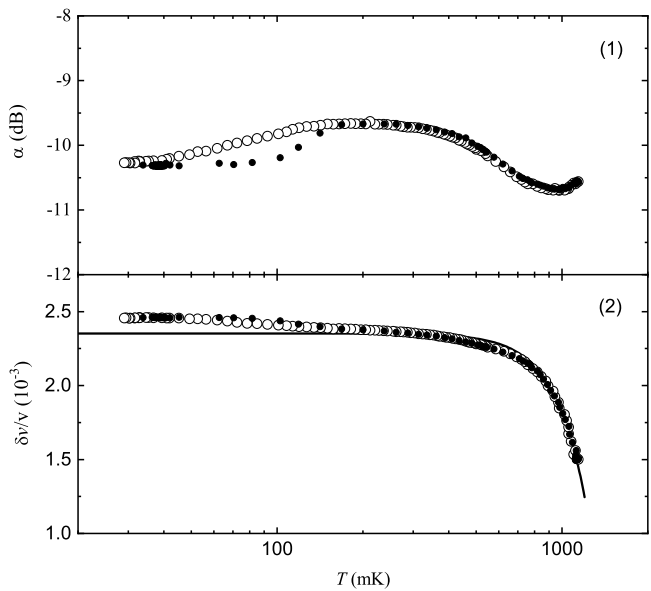}
\caption{
Attenuation (panel (1)) and change in speed (panel (2)) of a small-attenuation sample $\#4$ with 0.3 ppm of $^3$He impurity measured at $A_f = 2.51$ (black dots for warming and black open circles for cooling).  The number of plotted data is reduced.  The scale of ordinate in panel (1) is expanded compared with Fig. \ref{L1-2D-T-3drive-atten-dv-wc-v2} in order to show more detailed temperature dependence.  The $\delta v/v$ data during warming and cooling in panel (2) are almost indistinguishable and follows a $T^4$ dependence (black curve) expected from the phonon anharmonicity.
}
\label{D1b-atten-dv}  
\end{figure}

Three as-grown samples show a ``small-attenuation'' response over the measured temperature range as shown in Fig. \ref{D1b-atten-dv}.  The overall change in $\alpha$ is less than 3 dB and the hysteresis is small.  The data of $\delta v/v$ can be well described by a $T^4$ dependence as expected from the phonon anharmonicity\cite{Iwasa82} and there is no hysteresis between warming and cooling.  The small-attenuation response occurred in samples with both natural purity ($x_3$ = 0.3 ppm) and doped ($x_3$ = 20 ppm) samples.

\subsection{Anomalous sample}

\begin{figure}
\includegraphics[width=3.4in]{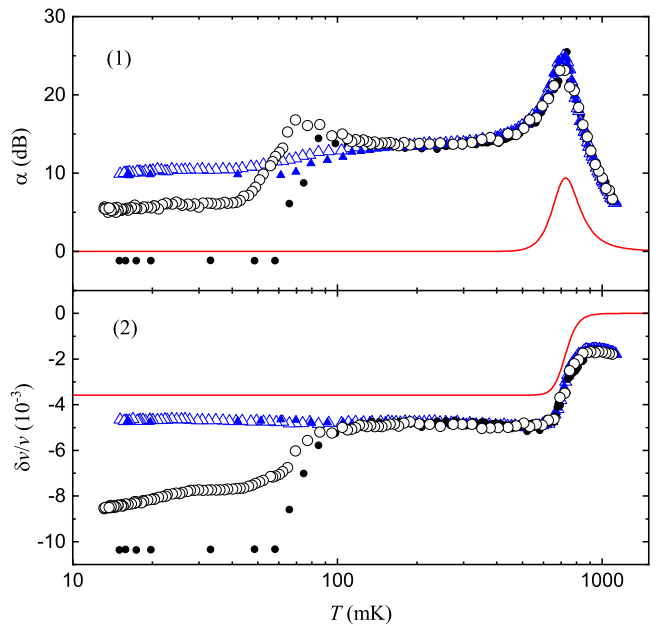}
\caption{
Attenuation (panel (1)) and change in speed (panel (2)) of an anomalous sample $\#6$ with $x_3$ = 0.3 ppm measured at $A_f = 1$ (black dots for warming(w) and black open cricles for cooling(c)) and $A_f = 2.82$ (blue filled(w) and open(c) triangles). The drive level was decreased to $A_i=0.45$ before starting the warming run at $A_f = 1$, while it was not changed at $A_f = 2.82$.  Red solid curves represent contributions from the Debye relaxation to the attenuation and speed (see Sec. \ref{anomalous discussion}).
}
\label{F1-attenuation-dv}
\end{figure}

The ultrasound behavior of the sample $\#6$ (a preliminary report\cite{Iwasa14} was given earlier) is unusual and anomalous.  The data from this sample taken at two drive levels $A_f$ = 1 and 2.82 are shown in Fig. \ref{F1-attenuation-dv}. Before starting the warming run at $A_f$ = 1, the drive level was decreased to $A_i$ = 0.45.  On the other hand, the drive level was not changed before starting the warming run at $A_f$ = 2.82.

In the high temperature range $T > 200$ mK in Fig. \ref{F1-attenuation-dv} both $\alpha$ and $\delta v/v$ are independent of the drive amplitude; the response is linear.  However, there is an anomalous peak in $\alpha$  and a rapid change in $\delta v/v$ around 700 mK. The behavior around 700 mK is totally unexpected from the GL theory.  It can be fitted as the Debye relaxation response (see Sec. \ref{anomalous discussion}).

In the low temperature range below 200 mK, the response shown in Fig. \ref{F1-attenuation-dv} is strongly sensitive to the drive amplitude.  The hysteresis of $\alpha$ at $A_f$ = 1 is similar to that of the typical sample $\#11$ at $A_f$ = 3.98 in Fig. \ref{L1-2D-T-3drive-atten-dv-wc-v2}.  (The critical amplitude in the sample $\#6$ seems $A_c<1$.)  The absence of thermal hysteresis at $A_f$ = 2.82 is due to a different measurement procedure from Fig. \ref{procedure} where the drive level was not decreased at $T_{\min}$ to a low $A_i$. 

\begin{figure}
\includegraphics[width=3.4in]{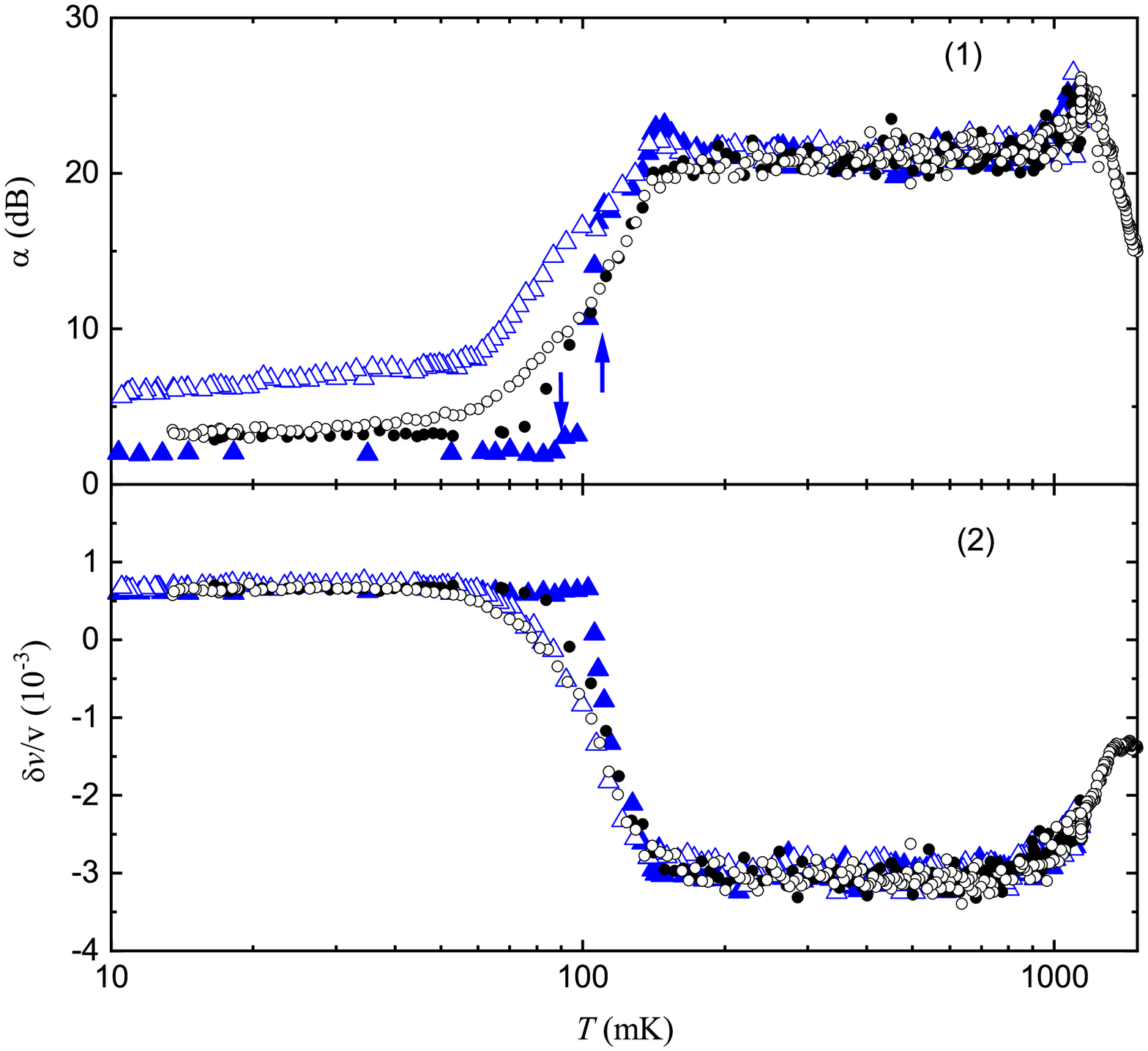}
\caption{
Attenuation (panel (1)) and change in speed (panel (2)) of the sample $\#8$ with $x_3$ = 20 ppm at $A_f = 1$ (black dots for warming(w) and black open cricles for cooling(c)) and $A_f = 1.41$ (blue filled(w) and open(c) triangles). The drive level was not changed before the warming run at $A_f = 1$, while it was increased from 1.0 to 1.41 at $A_f = 1.41$.  The down and up arrows in panel (1) indicate $T_1$ and $T_2$ at $A_f = 1.41$, respectively.
}
\label{H0e-atten-dv-38db}
\end{figure}

\subsection{Sample with $^3$He impurity concentration increased to 20 ppm}

To see if the anomalies in ultrasound propagation in solid $^4$He containing $x_3$ = 27.5 ppm impurity concentration observed by Ho et al.\cite{Ho97} could be reproduced in our experiment, $^3$He impurity concentration of 20 ppm was chosen.
The ultrasound response of the sample $\#8$ with $x_3 = 20$ ppm is shown in Fig. \ref{H0e-atten-dv-38db}.  The drive level was not changed before starting the warming run in the $A_f$ = 1.0 data while it was increased from 1.0 to 1.41 before the warming run in the $A_f$ = 1.41 data.  Probably this change in procedure is the reason why the initial value of $\alpha$ in the warming run at $A_f$ = 1.41 is smaller than that at $A_f$ = 1.0.  The critical amplitude in this sample seems $A_c<1$, like the case in the sample $\#6$.  The observed temperature dependence shows a similar attenuation peak as the anomalous sample $\#6$ but shifted to 1.2 K.  At high temperatures $T >$ 200 mK, the response is linear.  The response is nonlinear and hysteretic in the low temperature range similar to the typical sample response.  We estimate $T_1 = 80 \sim 90$ mK and $T_2 = 100 \sim 110$ mK from the warming runs.  Both of theses temperatures are higher than those for the $x_3 = 0.3$ ppm sample.

We note that $\delta v/v$ decreases with increasing temperature around 100 mK in this sample (Fig. \ref{H0e-atten-dv-38db}(2)) while it increases around 70 mK in the anomalous sample (Fig. \ref{F1-attenuation-dv}(2)).  The difference will be discussed in Sec. \ref{v-change}.

\section{Analysis}\label{Analysis}

In this section, the observed ultrasound response of the typical sample is analyzed.   The theory of dislocation lines by Granato and L\"{u}cke\cite{Granato56a}(GL) is introduced first. Pinning of dislocations by impurities is crucial in understanding the present experimental results at low temperatures.  The only significant impurities in our solid $^4$He samples are $^3$He atoms whose atomic volume is bigger than that of a $^4$He atom due to larger amplitude of zero-point vibration.  Thus $^3$He impurities can pin a dislocation line via elastic interaction.  The phenomena of pinning and unpinning in thermal equibrium as well as stress-induced unpinning are described.  Then, spatial variation of pinning is considered in order to explain the amplitude-dependent effects.  Based on these ideas, the data of $\alpha$ and $\delta v/v$ of the sample $\#11$ are analyzed.  Analyses and discussions of the samples with small attenuation, anomalous temperature dependence, and increased $^3$He impurity concentration will be deferred to Sec. \ref{Discussion}.

As the analysis involves nonlinearity, both the attenuation denoted by $\alpha$ and the attenuation coefficient $\beta$ are used.  In the linear case, they are simply related as $\alpha = \alpha_0 + \beta x_m$.  In the nonlinear case, on the other hand, spatial variation of the attenuation coefficent will be considered.

\subsection{Granato-L\"{u}cke theory}\label{G-L theory}\label{theory}
GL considered dislocation lines in the presence of ultrasound as vibrating strings that are pinned strongly at network nodes and weakly at impurities.  When ultrasound impinges on the dislocation lines, their displacements contribute an extra strain in addition to the elastic strain.  This interaction leads to changes in the propagation speed and the attenuation of the ultrasound\cite{Granato56a}.  Ultrasound in the MHz range excites resonant vibration in the dislocation lines.   We apply the GL theory to solid $^4$He from the overdamped vibration near 1.2 K down to underdamped vibration near 0.3 K.  Below 0.3 K, the impurities induce profound modifications on the ultrasound propagation.

The effect of dislocation lines on ultrasound response may be expressed\cite{Iwasa79} by writing the net fractional change in the longitudinal sound speed $v$ and the attenuation $\alpha$, respectively, as:
\begin{equation}
\frac{\delta v}{v} =  \frac{\delta v_d}{v} + \frac{\delta v_a}{v} + c,
\label{delta_v}
\end{equation}
and
\begin{equation}
\alpha = \alpha_d + \alpha_0,
\label{atten-tot}
\end{equation}
where $\delta v_a/v = a T^4$ accounts for the phonon anharmonicity($a$ is a fitting parameter), $\delta v_d/v$ and $\alpha_d$ represent the effects of the dislocation line motion on the speed and attenuation, respectively.   The constants $c$ and $\alpha_0$ are included as reference points which are to be determined in the data fitting procedure.  In the linear region, we can write 
\begin{equation}
\alpha_d = \beta_d x_m,
\label{linear_att}
\end{equation}
where $\beta_d$ is the attenuation coefficient due to dislocations.

According to GL, $\delta v_d/v$ and $\beta_d$ in the absence of impurity pinning are given by sums of the effects of all dislocation segments with a distribution in the network pinning length $L$:
\begin{equation}
\frac{\delta v_d}{v} =\frac{4Rv_t^2}{\pi^3 \Omega^2}\int_0^{\infty} \frac{L N(L)[1 - (\frac{L_0}{L})^2]dL}{[1 - (\frac{L_0}{L})^2]^2 + g^2},
\label{v_d}
\end{equation}
and
\begin{equation}
\beta_d=\frac{4Rv_t^2}{\pi^3 v\Omega}\int_0^{\infty}\frac{LN(L)g dL}{[1 - (\frac{L_0}{L})^2]^2 + g^2}.
\label{atten-dis}
\end{equation}
Here $R$ is the ``orientation factor" of the dislocation line relative to the sound propagation direction, $v_t$ the speed of transverse sound, and $g$ a dimensionless damping parameter.  The dislocation segments are assumed distributed in length by $N(L)dL$, where the distribution function $N(L)$ is the number of dislocation segments per unit volume per unit length.  The ``resonant length" $L_0$ is defined by:
\begin{equation}
L_0 = \sqrt{\frac{2}{1 - \nu}}\frac{v_t}{\Omega},
\label{L0}
\end{equation}
where $\nu (\approx 0.3)$ is the Poisson ratio.  Taking the speed of transverse sound at 35 bar as $v_t = 267$ m/s \cite{Franck70}, we estimate $L_0 = 7.5$ $\mu$m at the applied ultrasound frequency.  

The distribution function is not \emph{a priori} known. Typically\cite{Iwasa79}, the distribution of the network dislocation segments $L_n(L)$ is assumed to be a temperature independent exponential: 
\begin{equation}
N_n(L) = \frac{\Lambda}{L_{nA}^2}\exp(-\frac{L}{L_{nA}}),
\label{network-distribution}
\end{equation} 
where $\Lambda$ is the dislocation density, i. e. the total ``mobile" dislocation line length per unit volume and $L_{nA}$ is the average network pinning length.  Where the network dislocation segments dominate, $N(L)$ in Eq. (\ref{v_d}) and (\ref{atten-dis}) is replaced by $N_n(L)$.  Where temperature dependent impurity-pinned dislocation segments play crucial role in the ultrasound response, a new distribution function $N_A(L,T)$ will be introduced (see Eq. (\ref{network-impurity-distribution})).

The damping mechanism of dislocation line vibration in solid $^4$He has been identified experimentally\cite{Haziot13b} at low temperatures as the ``fluttering" interaction\cite{Ninomiya74} with phonons so that the temperature dependence of $g$ is given by $g = g_0 T^3$, where $g_0$ is a constant.  The dislocation line motion in our experiment varies over a wide range from the overdamped ($g > 1$) regime at high temperatures to underdamped ($g < 1$) regime at low temperatures.  The crossover from the overdamped to the underdamped vibration occurs around 560 mK where $g = 1$.

The major focus of this report is the underdamped regime of the dislocation line vibration at low temperatures, where the motion of those line segments of length close to $L_0$ results in ``resonant ultrasound response."   The integrand in Eq. (\ref{atten-dis}) shows that $\beta_d$ is characteristically dependent on the magnitude of $g$.  In the overdamped dislocation vibration regime, the denominator becomes approximately $g^2$ and $\beta_d$ is proportional to $g^{-1}$.  In the underdamped dislocation vibration regime, the integrand shows the resonant character around $L = L_0$. In the limit of $g << 1$, the height of the integrand is $L_0N(L_0)g^{-1}$ and the width is approximately $gL_0$.  Thus, $\beta_d$ in the underdamped regime at low temperatures becomes a constant independent of $T$: 

\begin{equation}
\beta_d(g<<1) \approx \frac{2R v_t^2L_0^2N_n(L_0)}{\pi^2 v \Omega}=\frac{2R\Lambda v_t^2L_0^2}{\pi^2 vL_{nA}^2 \Omega}\exp\left(-\frac{L_0}{L_{nA}}\right).
\label{atten-dis-lowT}
\end{equation}

The integrand in Eq. (\ref{v_d}) shows that $\delta v_d/v$ is determined by the balance between the negative change from those line segments with $L < L_0$ and positive change from those line segments with $L > L_0$.   The ``anti-resonance" character of the effects of dislocations on the propagation speed of ultrasound introduces some intricate dependence on temperature.  

\begin{table*}
\centering
\caption{
Parameters derived from the experimental data; the $^3$He concentration ($x_3$), the reference points for attenuation and speed of sound $(\alpha_0$, $c$), the anharmonic parameter ($a$), and the parameters related to dislocations ($g_0$, $R\Lambda$, $L_{nA}$, $L_{i0}$, and $E_b$).  The last column indicates the type of sample behavior.  The blank entries indicate that sufficient experimental data was not taken to enable extracting those parameters.
}
\begin{tabular}{ccccccccccc}
\hline\hline

Sample & $x_3$(ppm) & $\alpha_0$(dB) & $c$(10$^{-3}$) & $a$(10$^{-4}$K$^{-4}$) & $g_0$(K$^{-3}$) & $R\Lambda$(10$^9$ m$^{-2}$) & $L_{nA}$($\mu$m) & $L_{i0}(\mu$m) & $E_b$(K) & type \\ \hline
$\#1$ & 0.3 & -7.79 & 4.2 & -4 & 5.6 & 1.63 & 7.35 & 　 & 　 & typical \\ 
$\#3$ & 0.3 & -6.84 & -2.3 & -2.9 & 4.29 & 1.94 & 5.37 & 　 & 　  & typical \\ 
$\#4$ & 0.3 & -10.7 & 2.35 & -5.33 & 　 & 　 & 　 & 　 & 　& SA$^*$ \\ 
$\#6$ & 0.3 & -1.18 & -10.3 & -3.92 & 2.04 & 1.8 & 7.21 & 348 & 0.28 & anomalous \\ 
$\#7$ & 0.3 & -5.8 & 0.68 & -2.25 & 3.11 & 3.21 & 3.40 & 30 & 0.18 & typical  \\ 
$\#8$ & 20 & 2.0 & 0.60 &   & 　 & 　 & 　 & 　 & 　& anomalous \\ 
$\#9$ & 20 & -9.5 & 3.85 & -3.16 & 　 & 　 & 　 & 　 & 　& SA$^*$ \\ 
$\#10$ & 0.3 & 12.9 & -0.38 & -2.7 & 　 & 　 & 　 & 　 & 　 & typical  \\
$\#11$ & 0.3 & 10.31 & 0.9 & -4.87 & 5.63 & 1.71 & 3.22 & 101 & 0.35 & typical  \\ \hline\hline
\\
* small attenuation

\end{tabular}
\label{fit_parameters2}
\end{table*}

\subsection{Analysis of amplitude independent data}\label{LinearRegime}

\begin{figure}

\includegraphics[width=3.4in]{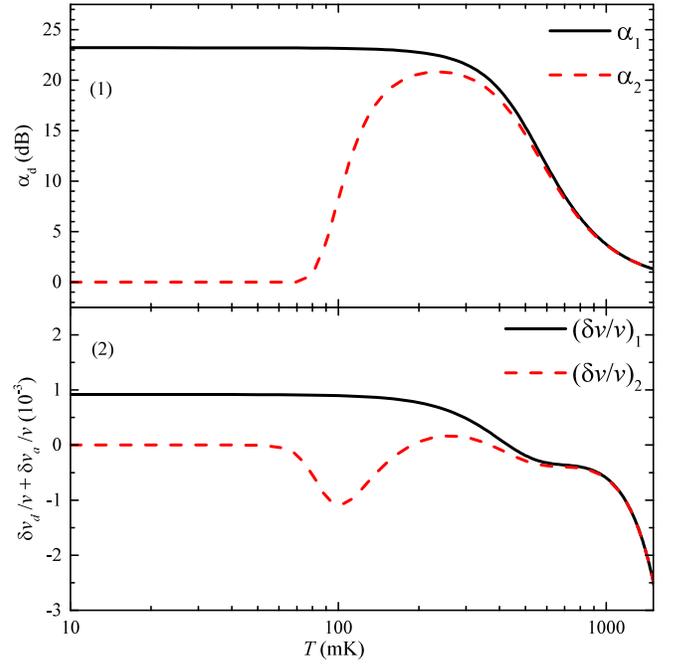}
\caption{
Calculated attenuation $\alpha_d$ (panel (1)) and change in speed $\delta v_d/v + \delta v_a/v$ (panel (2)) in the two limiting states of the sample  $\#11$.  The purely network-pinned state(black solid curves) are $\alpha_1$ and $(\delta v/v)_1$, and the purely impurity-pinned state(red dashed curves) are $\alpha_2$ and $(\delta v/v)_2$.
}
\label{atten-dv-pinned-unpinned-limits}
\end{figure}

At high temperatures, $T > 300$ mK, the impurities are thermally driven off of dislocations and pinning by impurities becomes negligible.  So the dislocations are pinned only at the network nodes.  The ultrasound response would then be independent of the drive amplitude as in fact observed above 300 mK in the typical sample $\#11$.  Consequently, the $\alpha$ data with $A_f = 14.1$ in Fig. \ref{L1-2D-T-3drive-atten-dv-wc-v2}(1) in the range $300 < T < 1150$ mK is fitted to Eq. (\ref{atten-tot}) with Eq. (\ref{linear_att}) and (\ref{atten-dis}) by assuming the exponential network distribution function Eq. (\ref{network-distribution}) and taking $\alpha_0$, $L_{nA}$, $R\Lambda$ and $g_0$ as adjustable parameters.   The best fit parameter values for the sample $\#11$ are listed in Table \ref{fit_parameters2}.   The  calculated $\alpha_d$ down to 10 mK with these fit parameters is shown as $\alpha_1$ in Fig. \ref{atten-dv-pinned-unpinned-limits}(1).  Note that $\alpha_1$ is essentially constant below 300 mK as expected from the underdamped motion of dislocation lines in the low temperature limit as given by Eq. (\ref{atten-dis-lowT}).

The $\delta v/v$ data with $A_f = 14.1$ in Fig. \ref{L1-2D-T-3drive-atten-dv-wc-v2}(2) in the range $400 < T < 1150$ mK is fitted to Eq.  (\ref{delta_v}) and (\ref{v_d}) by taking $a$ and $c$ as adjustable while fixing all other parameters, $L_{nA}$, $R\Lambda$ and $g_0$, listed in Table \ref{fit_parameters2} for the sample $\#11$ as determined above.  The resulting values of $a$ and $c$ are also listed in Table \ref{fit_parameters2}.  The calculated $\delta v_d/v + \delta v_a/v$ down to 10 mK with the parameters for the sample $\#11$ in Table \ref{fit_parameters2} is shown as $(\delta v/v)_1$ in Fig. \ref{atten-dv-pinned-unpinned-limits}(2).

The ``network-pinned" response, $\alpha_1$ and $(\delta v/v)_1$, reproduces the amplitude-independent data at $T > 300$ mK (see Fig. \ref{L1-3D-T-drive-atten} and \ref{L1-3D-T-drive-dv}) and represents the amplitude-dependent response at $T < 300$ mK in the limit of high amplitude.

\subsection{Pinning and unpinning of dislocations by impurities in thermal equilibirum}\label{ImpurityPinning}

The process of pinning and unpinning of dislocation lines by the $^3$He impurity atoms is central to understanding the amplitude-dependent and hysteretic ultrasound response observed at $T < 300$ mK.  In thermal equilibrium, the number of impurities on the dislocation lines is set by the balance between the temperature independent pinning rate and the thermally activated unpinning rate.\cite{Iwasa13a}. 

The average impurity pinning length in thermal equilibrium $L_{iA}(T)$ is expected to follow the Arrhenius law:
\begin{equation}
L_{iA}(T) = L_{i0}\exp\left(-\frac{E_b}{T}\right),
\label{impurity_length}
\end{equation}
where $E_b$ is the binding energy and $L_{i0}$ a constant.  
The temperature-dependent distribution function of the impurity pinning length is assumed to be exponential: 
\begin{equation}
 N_i(L,T) = \frac{\Lambda}{L_{iA}(T)^2} \exp\left(-\frac{L}{L_{iA}(T)}\right). 
\label{impurity-distribution}
\end{equation}
Then the effective temperature-dependent distribution function of the combined impurity-network pinning length is also exponential:
\begin{equation}
 N_A(L,T) = \frac{\Lambda}{L_A(T)^2} \exp\left(-\frac{L}{L_A(T)}\right),
\label{network-impurity-distribution}
\end{equation}
where the effective average pinning length $L_A(T)$ is taken as the parallel combination of $L_{nA}$ and $L_{iA}(T)$:
\begin{equation}
L_A(T) = \frac{L_{nA}L_{iA}(T)}{L_{nA} + L_{iA}(T)}.
\label{impurity_network_length}
\end{equation}

\begin{figure}
\includegraphics[width=3.4in]{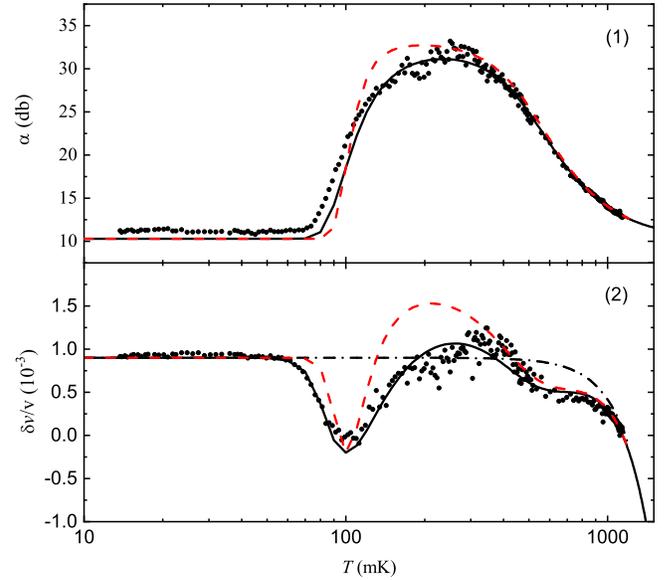}
\caption{
Cooling data of the sample $\#11$ at $A_f$ =  1.12 (black closed circles) compared with calculated curves.  Panels (1) and (2) show attenuation and change in speed, respectively.  Black solid curves represent calculations with $E_b=0.35$ K and other parameters shown in Table \ref{fit_parameters2} for sample $\#11$.  They are the same curves as $\alpha_2$ and $(\delta v/v)_2$ in Fig. \ref{atten-dv-pinned-unpinned-limits} shifted by $\alpha_0=10.31$ dB and $c=0.9\times 10^{-3}$.  Red broken curves represent calculations with $E_b=0.67$ K, $L_{i0}=2440$ $\mu$m and other parameters in Table \ref{fit_parameters2} for sample $\#11$.  Black dash-dotted curve in panel (2) represents the contribution of the phonon anharmonicity to the change in speed.
}
\label{binding_energy}
\end{figure}

When the ultrasound drive amplitude is sufficiently small, the dislocations are pinned at network nodes and by $^3$He impurities in thermal equilibrium.  This pinning state will be simply called as “impurity-pinned” state.  The $\alpha$ and $\delta v/v$ data at $A_f = 1.12$ shown in Fig. \ref{L1-2D-T-3drive-atten-dv-wc-v2}(1) and (2), respectively, are established as the impurity-pinned state of dislocations based on the observations that the data from warming and cooling runs are identical and $\alpha$ is the lowest among the runs.  In this case, we assign $N_A(L,T)$ as the distribution function $N(L)$ in Eq. (\ref{v_d}) and (\ref{atten-dis}).  The attenuation data of the cooling run with $A_f = 1.12$ in the entire temperature range is fitted with Eq. (\ref{atten-dis}) (together with Eq. (\ref{impurity_length}), (\ref{network-impurity-distribution}), and (\ref{impurity_network_length})) with two fitting parameters $L_{i0}$ and $E_b$ while keeping $\alpha_0$, $g_0$, $R\Lambda$, and $L_{nA}$ fixed to those found in Sec. \ref{LinearRegime} and listed in Table \ref{fit_parameters2} for the sample $\#11$. The best fitting parameters are $L_{i0}=101$ $\mu$m and $E_b=0.35$ K as shown in Table \ref{fit_parameters2}.  With these parameters, $\alpha_d$ and $\delta v_d/v + \delta v_a/v$ are calculated and shown in Fig. \ref{atten-dv-pinned-unpinned-limits} as $\alpha_2$ and $(\delta v/v)_2$, respectively.  They represent the response of the impurity-pinned state in the limit of low amplitude.
Figure \ref{binding_energy} shows the measured attanuation and speed in sample $\#11$ in the low amplitude limit at $A=1.12$ and the calculated curves for the impurity-pinned state along with other curves for $E_b=0.67$ K(see Sec. \ref{ShearModulus}) and $\delta v_a/v+c$.

\begin{figure}
\includegraphics[width=3.4in]{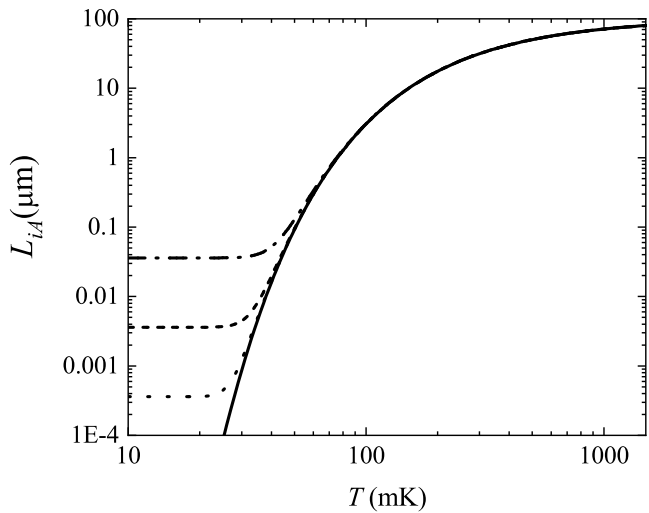}
\caption{
Temperature dependences of the average impurity pinning length calculated from Eq. (\ref{impurity_length3}) for $x_3$ = 0.3 ppm with $L_{i0} = 101$ $\mu$m and $E_b$ = 0.35 K for $\gamma$ = 0 (solid line), 1 (dotted line), 10 (dashed line), and 100 (dash-dotted line).
}
\label{impurity-length}
\end{figure}

The Arrhenius law,  Eq. (\ref{impurity_length}), becomes unphysical at low temperatures around $T_{\min}$ because the minimum value of the impurity pinning length is limited to the nearest-neighbor atomic distance, $b = 0.36$ nm.  A more realistic temperature dependence would be
\begin{equation}
L_{iA}(T) = L_{i0}\exp\left(-\frac{E_b}{T}\right) + \gamma b, 
\label{impurity_length3}
\end{equation}
where $\gamma$ is a numerical factor. Setting $\gamma$ = 0, Eq. (\ref{impurity_length3}) is equivalent to Eq. (\ref{impurity_length}).  We expect $\gamma>1$ because of the repulsive force between neighboring $^3$He atoms on a dislocation line.  Figure \ref{impurity-length} shows the temperature variation of $L_{iA}$ for $\gamma$ = 0, 1, 10, and 100.
The lines of $\alpha_2$ and $(\delta v/v)_2$ in Fig. \ref{atten-dv-pinned-unpinned-limits} are calculated with $\gamma$ = 0, but even when $\gamma$ = 100, the deviations from these lines are insignificant.  Thus the value of $\gamma$ cannot be determined from the present measurements.

Let us consider the variation of $L_{iA}$ with $x_3$.  For simplicity, we assume that the number of $^3$He atoms contributing to pinning the dislocation lines is proportional to $x_3$.  Then we expect 
\begin{equation}
L_{iA}(x_3,T) = \frac{K}{x_3}\exp\left(-\frac{E_b}{T}\right),
\label{impurity_length2}
\end{equation}
from Eq. (\ref{impurity_length}) where $K/x_3=L_{i0}$.  We obtain $K=30$ $\mu$m$\cdot$ppm from assuming $x_3$=0.3 ppm and $L_{i0}=101$ $\mu$m.

\subsection{Stress-induced unpinning}\label{catastrophic}
As noted in Sec. \ref{theory}, the dislocation lines are strongly pinned at network nodes but also weakly pinned by impurities.  As a $^3$He atom has a larger atomic volume than a $^4$He atom, the difference in the atomic size causes impurity pinning.  The interaction between a dislocation and a $^3$He atom can be described by the elastic potential energy (see Appendix \ref{AppendixC}, Eq. (\ref{C-1})).  If the applied force $F$ on a pinning impurity is sufficiently large, the dislocation line can be unpinned there.  The compressional stress of the longitudinal ultrasound induces shear stress component $\sigma_s$ on the glide plane of a dislocation line and $\sigma_s$ in turn induces a force on the dislocation line.  When the dislocation segments with lengths $L_1$ and $L_2$ adjoining a pinning site bow out, the unpinning force on the pinning site is given by
\begin{equation}
F = b\sigma_s \frac{L_1+L_2}{2},
\label{force-equation}
\end{equation}
where $b$ is the magnitude of Burgers vector which is equal to the nearest neighbor atomic distance. A critical force, $F_c$, is required to produce unpinning. The condition of unpinning, $F>F_c$, can be transformed to  
\begin{equation}
L_1 + L_2 > L_c ,
\label{Lc-equation-2}
\end{equation}
where $L_c$ is the critical length\cite{Iwasa13} 
\begin{equation}
L_c = \frac{2 F_c}{b\sigma_s}.
\label{Lc-equation}
\end{equation}
For our sample $\#11$, $L_c$ is estimated to be (see Appendix \ref{AppendixB}, Eq. (\ref{B-9}))
\begin{equation}
L_c [\mu \mbox{m}] = \frac{12}{A}.
\label{Lc-numerical}
\end{equation}
The critical force is estimated (see Appendix \ref{AppendixB}) to be $F_c=1.5\times 10^{-15}$ N.  For comparison, $F_c$ has been evaluated\cite{Iwasa13} from a torsional-oscillator experiment\cite{Aoki07} to be $1.0\times 10^{-16}$ N  and from a shear-modulus experiment\cite{Fefferman14} to be $6.8\times 10^{-15}$ N.

There are two types of stress-induced unpinning in our experiment.  One is the ``catastrophic"\cite{Granato56a} unpinning which occurs when the stress is increased at low temperatures.  The other is unpinning of a single impurity atom on a network segment which occurs when the sample is cooled at constant stress amplitude.  The catastrophic unpinning is discussed first.

Consider a network segment, i. e. a dislocation line pinned by two adjacent network nodes, which is additionally pinned by several impurities distributed between the two ends under a small external stress.  When the external shear stress increases as the ultrasound drive amplitude is increased, $L_c$ decreases according to Eq. (\ref{Lc-numerical}).  Eventually, $L_c$ becomes shorter than the sum of some adjacent two impurity segments along the dislocation line and the dislocation is unpinned from the impurity.  Once unpinning is initiated, it continues catastrophically to the next adjacent impurity segments till this network segment is pinned only at its two end network nodes.

In the original model of the catastrophic unpinning\cite{Granato56a}, the pinning impurities are assumed to be immobile so that the dislocation line is pinned by the impurities again as soon as the stress is decreased.  In the case of solid helium, on the other hand, the $^3$He impurities drift away from the dislocation line once unpinned.  Therefore, as the catastrophic unpinning is initiated, the network segment becomes free from impurity pinning.

Consider next the situation when a network segment of length $L_N$ is pinned by a single impurity atom, which divides the network segment into two impurity segments of lenghts $L_1$ and $L_2$ where $L_1 + L_2 = L_N$.  When an ultrasound pulse at a drive level A corresponding to the critical length $L_c$ is applied, the impurity atom is unpinned if $L_N > L_c$, while it remains pinned if $L_N < L_c$.

In a cooling run, pinning by impurities becomes appreciable at temperatures below $E_B$ ($\approx 350$ mK).  According to the calculation in Fig. \ref{impurity-length}, $L_{iA}$ at 300, 200, and 100 mK are 31.2, 17.4, and 3.0 $\mu$m, respectively.  For example, a network segment with $L_N \approx 3$ $\mu$m is pinned by one $^3$He impurity atom on average when the sample is cooled down to 100 mK.  If the amplitude of ultrasound is $A$ = 4 corresponding to $L_c = 3$ $\mu$m, longer network segments ($L_N > 3$ $\mu$m) pinned by an impurity are unpinned because $L_N > L_c$, while shorter ones ($L_N < 3$ $\mu$m) pinned by an impurity remain pinned.  When the sample is further cooled, the dynamical aspects of pinning and unpinning must be taken into account.\cite{Iwasa13}  The temperature-independent pinning rate on a network segment of length $L_N = 3$ $\mu$m is about 1 atom/s (see Eq. (\ref{R1})), which is much smaller than the pulse repetition frequency (1 kHz).
When a $^3$He atom happens to approach a network segment of length $L_N > L_c$ and pins it down just after an ultrasound pulse has traversed, the next ultrasound pulse comes within 1 ms and unpins the network segment from the impurity atom.  As a result, the network segments longer than $L_c$ stay unpinned down to $T_{\min}$.  Those shorter than $L_c$, on the other hand, are pinned by impurities and the average pinning length decreases with decreasing temperature according to Eq. (\ref{impurity_network_length}). 

\begin{figure}
\includegraphics[width=3.4in]{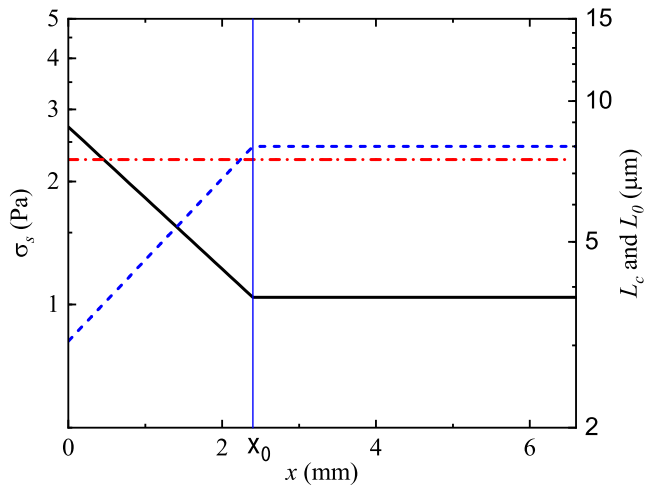}
\caption{
Schematic spatial variations of shear stress amplitude and critical length at $T_{\min}$.  The driver transducer is located at $x = 0$ and the receiver transducer at $x_m = 6.6$ mm.  The drive amplitude is taken as $A_f = 4$, which leads to $\sigma_{s0} = 2.72$ Pa from the transducer calibration constant(see Eq. (\ref{A-2})).  The shear stress amplitude (solid black line) is assumed to vary near the drive transducer according to $\sigma_s(x) = \sigma_{s0}\exp(-\beta x)$, where $\beta = 380$ m$^{-1}$ is the attenuation coefficient in the network-pinned region. The sample divides into two regions separated by the border at $x_0 = 2.4$ mm; the network-pinned region ($x < x_0$) where $L_c$ (dashed blue line) is shorter than the resonance length $L_0$ (dash-dotted red line) and the impurity-pinned region ($x > x_0$) where $L_c \geq L_0$.  The critical force $F_c = 1.5\times 10^{-15}$ N is assumed.
}
\label{stress-Lc-vs-x}

\end{figure}

\subsection{Spatial variation of pinning}\label{spatial variation}

In the previous section, the ultrasound is assumed to travel unattenuated throughout the sample.  We introduce in this section a more realistic model that the amplitude of the ultrasound pulse is attenuated along the path through the sample.  The local amplitude is written as $A(x)$ where $x$ is the position in the sample from the driver transducer end ($x=0$ mm) towards the receiver end ($x_m=6.6$ mm).  Similarly the amplitude of local shear stress and the local critical length are given by $\sigma_s(x)$ and $L_c(x)$, respectively.
In order to analyze the amplitude dependent attenuation below 300 mK, the spatial variation of the response of dislocation lines to ultrasound field must be considered.  In the underdamped regime of dislocation vibration, the network segments of lengths around $L_0$ mainly contribute to $\alpha_d$ and $\delta v_d/v$ owing to their resonant behavior.  As the attenuation in the sample $\#11$ in the network-pinned state at $T<300$ mK is typically 20 dB, $A(x)$ and $\sigma_s(x)$ decrease by a factor of 10 while $L_c(x)$ increases by a factor of 10 within the sample from $x$ = 0 to $x_m$.  An example of spatial variations of $\sigma_s(x)$ and $L_c(x)$ with the drive amplitude initially set to $A_f = 4.0$, corresponding to $L_c = 3.0$ $\mu$m at $x = 0$ mm, is illustrated in  Fig. \ref{stress-Lc-vs-x}.

The local $L_c(x)$ increases with $x$ and eventually exceeds $L_0 = 7.5$ $\mu$m.  In Fig. \ref{stress-Lc-vs-x}, $x_0$ is the position where $L_c$ becomes equal to $L_0$.  The sample can be divided into two regions of distinct ultrasound response to dislocations: (1)``network-pinned" state region nearer to the driver, $0 < x < x_0$, and (2)``impurity-pinned" state region nearer to the receiver, $x_0 < x < x_m$.  We define a characteristic fraction $r$ as $r=x_0/x_m$. ($r = 0.36$ in Fig. \ref{stress-Lc-vs-x}).

In the network-pinned region, the network segments with lengths near $L_0$ are not at all pinned by impurities.  This region corresponds to the high drive amplitude limit considered in Sec. \ref{LinearRegime}.  The ultrasound response is thus described by $\alpha_1$ and $(\delta v/v)_1$ shown in Fig. \ref{atten-dv-pinned-unpinned-limits}.

In the impurity-pinned region, $L_c$ is longer than $L_0$ so that the network segments with lengths near $L_0$ are pinned by $^3$He impurities.  This region corresponds to the low drive amplitude limit considered in Sec. \ref{ImpurityPinning}.  The impurity pinning length given by Eq. (\ref{impurity_length}) is strongly temperature dependent and the ultrasound response is thus approximately described by $\alpha_2$ and $(\delta v/v)_2$ shown in Fig. \ref{atten-dv-pinned-unpinned-limits}.
The ultrasound attenuation $\alpha_2$ becomes vanishingly small at low temperatures below 70 mK, so that $\sigma_s$ and $L_c$ are spatially independent at $T_{\min}$ in the impurity-pinned region ($x>x_0$) in Fig. \ref{stress-Lc-vs-x}.

The net ultrasound response of attenuation and change in speed to the dislocation motion is written as:
\begin{eqnarray}
\alpha_d = r\alpha_1 + (1-r)\alpha_2
\label{r_atten_fit}\\
\frac{\delta v_d}{v} = r\left(\frac{\delta v}{v}\right)_1 + (1-r)\left(\frac{\delta v}{v}\right)_2 .
\label{r_dv_fit}
\end{eqnarray}
The distinction of network-pinned and impuritiy-pinned regions becomes meaningless at $T > 300$ mK as $L_{iA}$ becomes longer than $L_0$.  Nevertheless, Eq. (\ref{r_atten_fit}) and (\ref{r_dv_fit}) can be applied in the whole temperature range for convenience since $\alpha_d$ and $\delta v_d/v$ are independent of $r$ at $T > 300$ mK.

\subsection{Analysis of cooling runs at various $A_f$} \label{analysis-cooling}

\begin{figure*}
\includegraphics[width=7.0in]{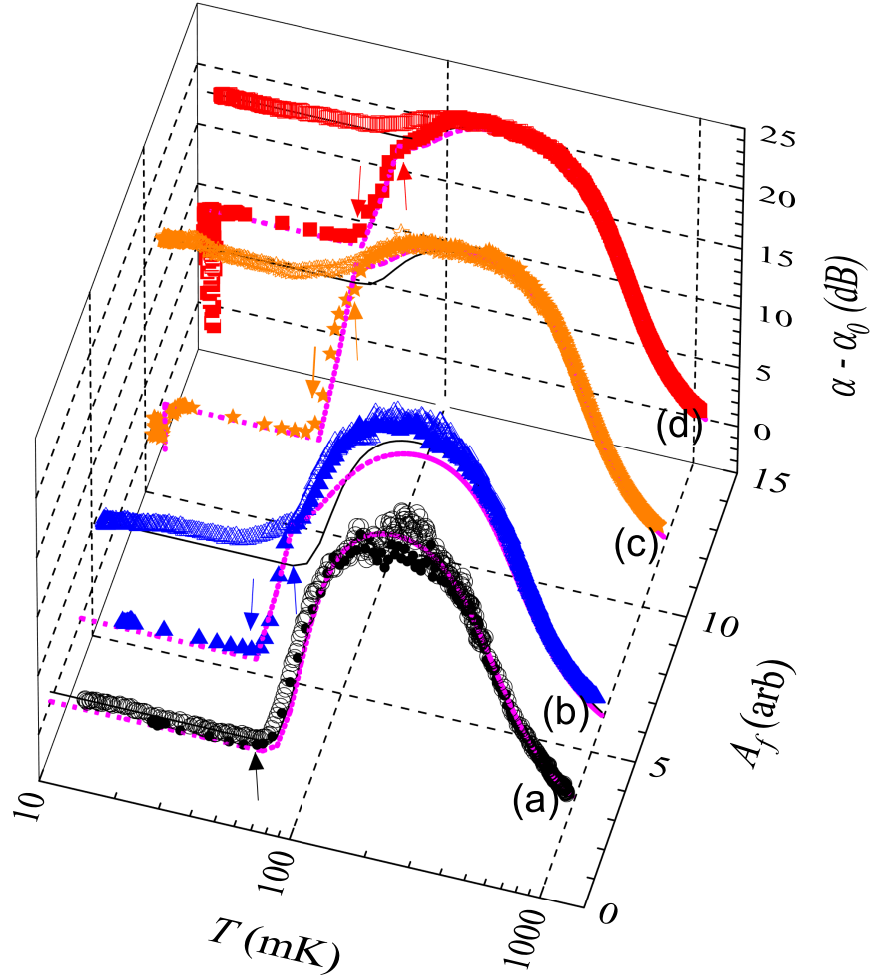}
\caption{
Temperature dependence of $\alpha - \alpha_0$ in the sample $\#11$ during warming(closed symbols) and cooling(open symbols) when $A_f$ =  1.12(a), 3.98(b), 10.0(c) and 14.1(d).  Data and symbols are identical to those in Fig. \ref{L1-2D-T-3drive-atten-dv-wc-v2}(1) except the added warming data (closed orange stars) and cooling data (open orange stars) for (c).  Curves(black solid for cooling and dotted magenta for warming) are $\alpha_d$ calculated from Eq. (\ref{r_atten_fit}) with the values of $r$ in Fig. \ref{r-warming-cooling}.  Down(up) arrows indicate characteristic temperatures $T_1$($T_2$) where the dislocation line pinning state changes.  Vertical data at $T_{\min}$ in the warming runs of (c) and (d) indicate time dependent response as in Fig.\ref{L1-2D-T-3drive-atten-dv-wc-v2}.
}
\label{L1-3D-T-drive-atten}
\end{figure*}

\begin{figure*}
\includegraphics[width=7.0in]{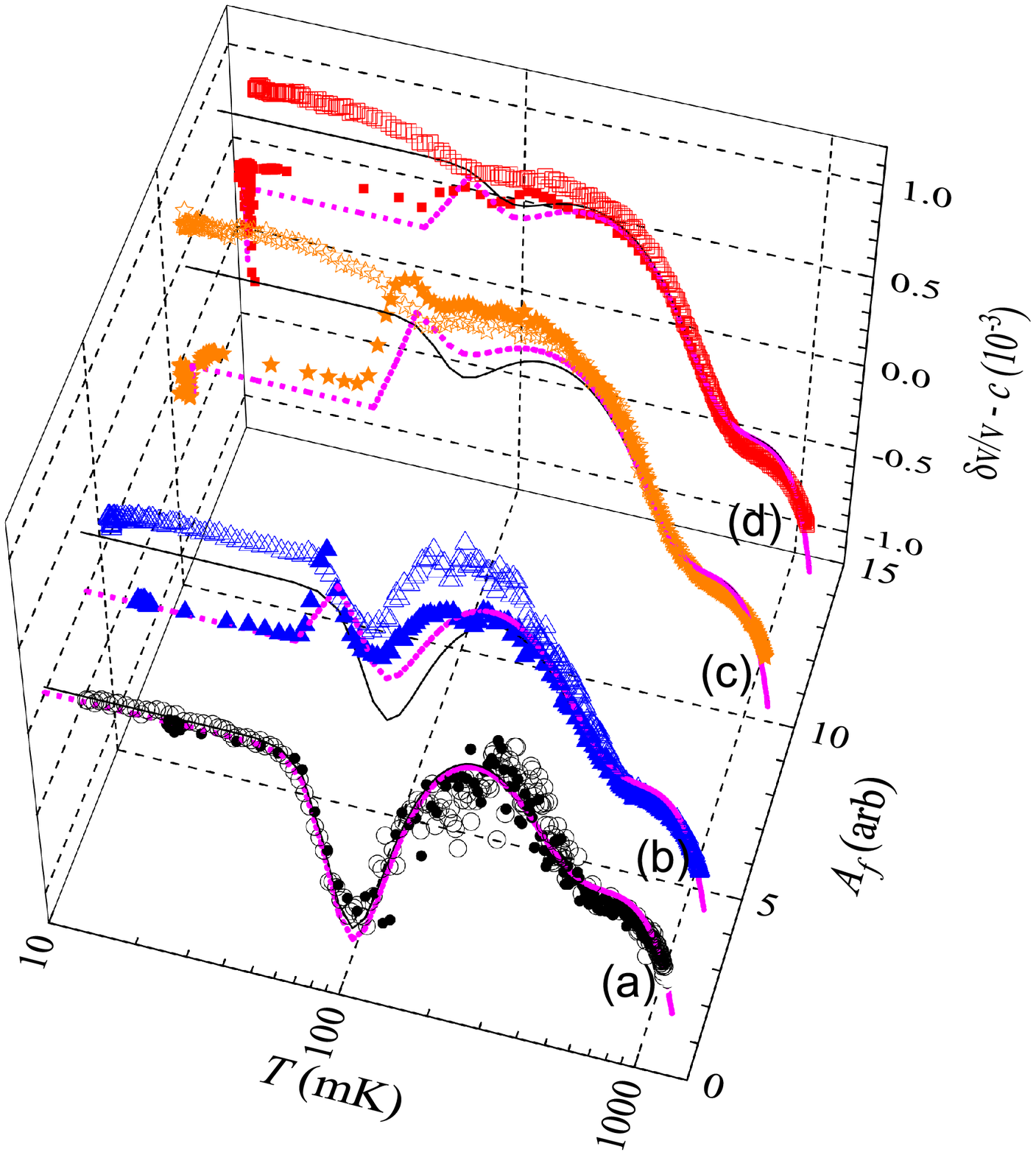}
\caption{
Temperature dependence of $\delta v/v - c$ in the sample $\#11$ when $A_f$ =  1.12(a), 3.98(b), 10.0(c) and 14.1(d).  The meanings of symbols are the same as in Fig. \ref{L1-3D-T-drive-atten}.  Curves are $\delta v_d/v + \delta v_a/v$ calculated from Eq. (\ref{r_dv_fit}) with the values of $r$ in Fig. \ref{r-warming-cooling}.  Vertical data at $T_{\min}$ in the warming runs of (c) and (d) indicate time dependent response.
}
\label{L1-3D-T-drive-dv}
\end{figure*}

To facilitate the analysis of data with Eqs. (\ref{r_atten_fit}) and (\ref{r_dv_fit}), the reference points $\alpha_0$ and $c$ are subtracted from the experimental data of the sample $\#11$ shown in Fig. \ref{L1-2D-T-3drive-atten-dv-wc-v2}.  The shifted data together with an additional data set with $A_f = 10.0$ are plotted in Fig. \ref{L1-3D-T-drive-atten} and \ref{L1-3D-T-drive-dv} for the attenuation and change in speed, respectively.

The characteristic fraction $r$ is assumed to be a temperature-independent constant in each cooling run because $\alpha_1$ is almost constant below 300 mK.  As $\alpha_2 = 0$ at $T_{\min}$, the value of $r(A_f)$ is determined from Eq. (\ref{r_atten_fit}) as
\begin{equation}
r(A_f) = \frac{\alpha(A_f) - \alpha_0}{\alpha_1 (T_{\min})} ,
\label{r-cooling}
\end{equation}
where $\alpha(A_f) - \alpha_0$ is the attenuation of a cooling run at $T = T_{\min}$ in Fig. \ref{L1-3D-T-drive-atten}.

\begin{figure}
\includegraphics[width=3.4in]{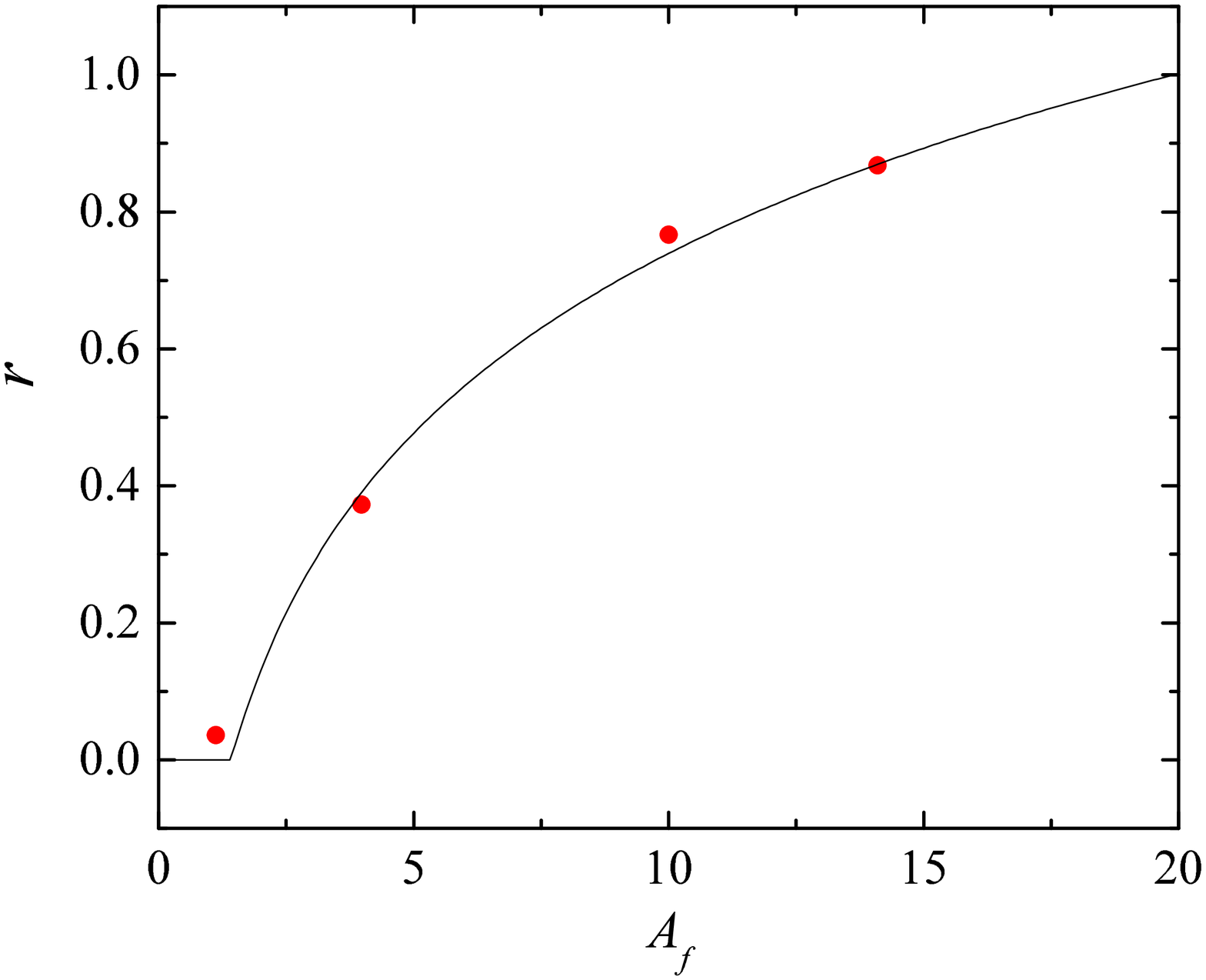}
\caption{
Empirical and fitted values of r for the cooling runs of the sample $\#11$ are plotted versus  $A_f$ as red filled circles and a black line, respectively.
}
\label{r-Af}
\end{figure}

The empirical values of $r(A_f)$ for the cooling runs with $A_f =$ 1.12, 3.98, 10.0, and 14.1 are 0.036, 0.373, 0.767, and 0.868, respectively, as shown in Fig. \ref{r-Af} and in Fig. \ref{r-warming-cooling}(2).  Fitting of the data except for $A_f =$ 1.12 results in
\begin{equation}
r(A_f) = \begin{cases}
	0 & (0 \leq A_f < 1.54)	\\
	\frac{1}{2.5} \ln \frac{A_f}{1.54} & (1.54 \leq A_f)
	\end{cases}
\label{r-theory}
\end{equation}
which is also shown in Fig. \ref{r-Af}.  The functional form of Eq. (\ref{r-theory}) is derived in Appendix \ref{AppendixB}.

The values of $r(A_f)$ in cooling runs are entered in Eq. (\ref{r_atten_fit}) to calculate $\alpha_d$.  The resulting fits are drawn as black solid lines in Fig. \ref{L1-3D-T-drive-atten}.  The observed temperature dependence of attenuation is generally well reproduced by the fits.
The same numerical values are used in Eq. (\ref{r_dv_fit}) together with the phonon anharmonicity term, $\delta v_a/v $, to calculate the fit curves to the $\delta v/v - c$ data of the cooling runs shown in Fig. \ref{L1-3D-T-drive-dv} as black solid lines.  The agreement between the experimental and calculated changes in speed at $T_{\min}$ is not as good as that of attenuation likely because $r$ is determined from the attenuation data alone.

\subsection{Analysis of warming runs} \label{analysis-warming}

According to the measurement procedure in Fig. \ref{procedure}, the drive amplitude is decreased to $A_i (< 0.07)$ corresponding to $L_c > 170$ $\mu$m at $T_{\min}$ as estimated from Eq. (\ref{Lc-numerical}) before starting each warming run.  Practically all the network segments are densely pinned by $^3$He impurity atoms so that each warming run starts with the totally impurity-pinned state ($r = 0$).  The drive level is then increased to a new $A_f$ (see [5]$\rightarrow$[1]$\rightarrow$[2] in Fig. \ref{procedure}). 

When $A_f = 1.12$, the critical length is estimated to be $L_c = 10.7$ $\mu$m so that stress-induced unpinning does not occur for the network segments with lengths near $L_0 = 7.5 $ $\mu$m (see Fig. \ref{procedure}(2a) and (3a)).  Therefore we expect $r = 0$ throughout the warming run as shown in Fig. \ref{r-warming-cooling}(1).  The calculated curves of $\alpha_d$ and $\delta v_d/v + \delta v_a/v$ for the warming run at $A_f = 1.12$ (dotted magenta lines in Fig. \ref{L1-3D-T-drive-atten} and \ref{L1-3D-T-drive-dv}, respectively) coincide with those for the cooling run.
The increase in $\alpha_d$ at temperatures above $T_2$ is due to the thermal unpinning effect described by Eq. (\ref{impurity_length}).

When $A_f = 3.98$, $L_c$ is decreased to 3.0 $\mu$m.  Stress-induced unpinning, however, does not occur at $T_{\min}$ because the average impurity pinning length, $L_{iA}(T_{\min})$, is much shorter than this $L_c$ (see Fig. \ref{procedure}(2b)).  As the temperature is increased, $L_{iA}(T)$ becomes longer.  At  $T = T_1  (\approx 50$ mK), the condition of unpinning, Eq. (\ref{Lc-equation-2}), is evidently satisfied for some inpurity-pinned segments and the catastrophic unpinning is initiated (see Fig. \ref{procedure}(3b)) so that $r$ starts to increase as shown in Fig. \ref{r-warming-cooling}(1).
More network segments are catastrophically unpinned and $r$ continues to increase as the temperature is further increased.  At the same time, however, the relative ultrasound amplitude at the border between the network- and impurity-pinned regions, $A(x_0)$, decreases due to the attenuation in the extended network-pinned region.  At $T = T_2 (\approx 70$ mK), $A(x_0)$ is expected to become smaller than 1.54 and $L_c$ becomes longer than $L_0$.  Indeed the value of $r$ at 70 mK is estimated to be 0.5, from which we obtain $A(x_0)$ = 1.26.  A closer inspection shows that there is a kink in the attenuation data at $T_2$.  We assume that $r$ stays constant at $T > T_2$ as shown in Fig. \ref{r-warming-cooling}(1).

When the drive amplitude is increased to $A_f = 10.0$ and 14.4, catastrophic unpinning evidently already starts at $T_{\min}$ (see Fig. \ref{procedure}(2c)).  As a result, $\alpha$ increases (with accompanied time dependence, see Fig. \ref{amp-dv-relaxation}(3)) at $T_{\min}$ as shown in Fig.\ref{L1-3D-T-drive-atten}(c) and (d).  In terms of $r$, it increases from $r = 0$ to 0.17 and 0.44 for $A_f = 10.0$ and 14.4, respectively, as shown in Fig. \ref{r-warming-cooling}(1) and the relative ultrasound amplitude at the border are estimated to be $A(x_0) = $6.53 and 4.79, respectively.
When the sample is warmed, thermally-assisted stress-induced catastrophic unpinning occurs between $T_1$ and $T_2$ similar to the warming run at $A_f = 3.98$.  The experimental values of $r$ at $T_2$ are 0.83 for both $A_f = 10.0$ and 14.4, and $A(x_0)$ decrease to 1.25 and 1.81 for $A_f = 10.0$ and 14.4, respectively.

\subsection{Extraction of parameters and consistent description of ultrasound response}\label{extract-parameters}

\begin{figure}
\includegraphics[width=3.4in]{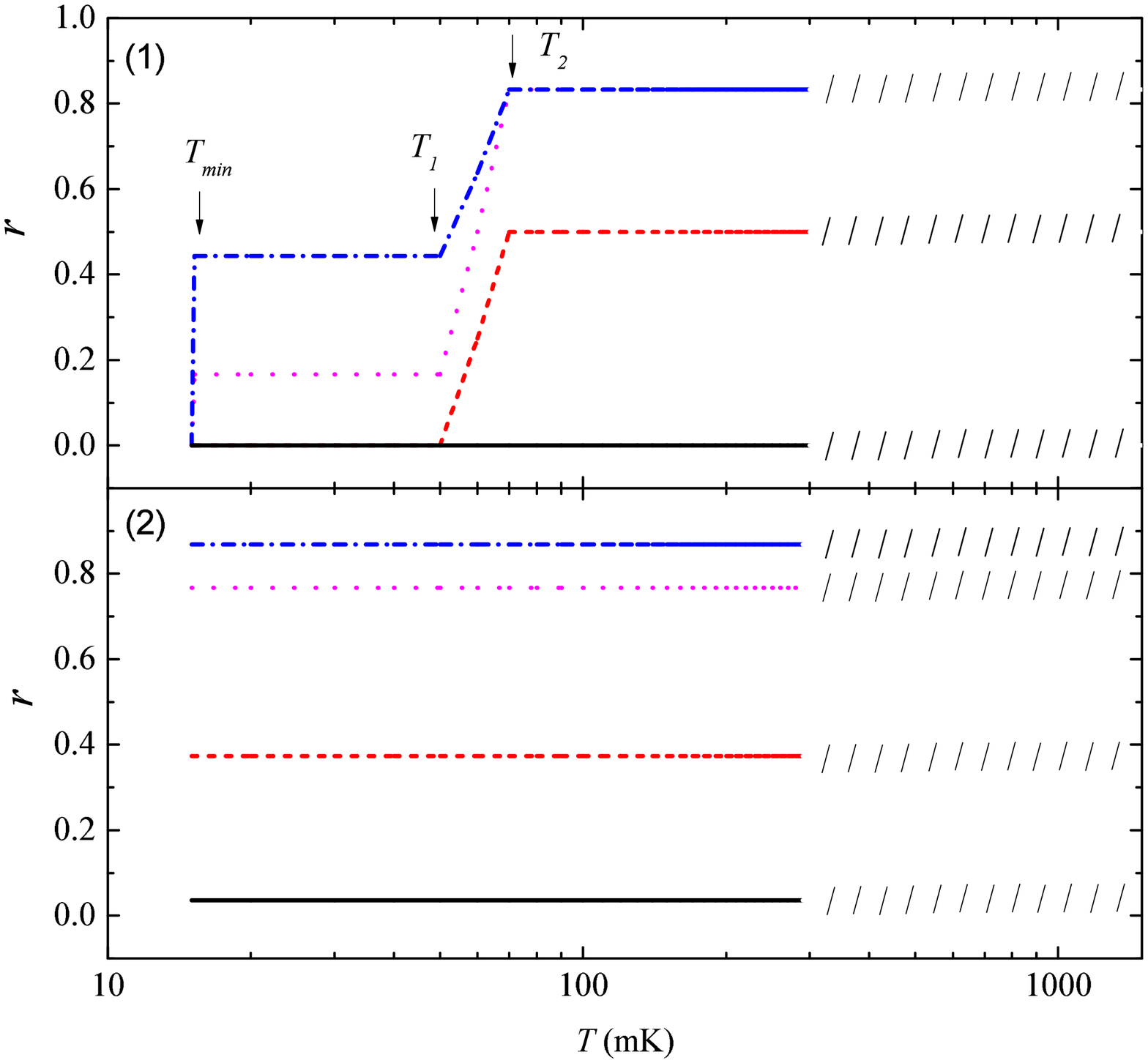}
\caption{
Dependence of $r$ on $A_f$ and $T$ during warming(1) (Sec. \ref{analysis-warming}) and cooling(2) (Sec. \ref{analysis-cooling}) runs: $A_f$ =  1.12(continuous(black line)), 3.98(dashed(red)), 10.0(dotted(magenta)) and 14.1(dash-dotted(blue)).  In the warming runs, the drive amplitude is increased from $A_i$ to $A_f$ at $T_{\min}$, thermally-assisted stress-induced unpinning of impurity sites occurs where $T_{1} < T < T_{2}$, and only thermal unpinning occurs where $T > T_{2}$.  In the cooling runs, the values of $r$ do not vary with $T$.  The slanted hash marks indicate the temperature range where the calculated ultrasound response does not depend on $r$.
}
\label{r-warming-cooling}
\end{figure}

The calculated attenuation and speed using Eq. (\ref{r_atten_fit}), (\ref{r_dv_fit}) and the temperature dependence of $r$ shown in Fig. \ref{r-warming-cooling} are drawn as dotted lines for warming runs and solid lines for cooling runs in Fig. \ref{L1-3D-T-drive-atten} and \ref{L1-3D-T-drive-dv}.  The observed and calculated attenuation agree well except in the cooling runs from 350 mK to $T_1$.  The cause of dicrepancy may be the oversimplified assumption of a constant value of $r$ for each cooling run.  The agreement between the observed and the calculated speed of sound is worse than that of attenuation likely because $r$ is determined only from the attenuation data.  Nevertheless, the local ``peaks" and ``valleys" found in the observed temperature dependence of $\delta v/v$ are reproduced by the calculations.

The hysteretic ultrasound response versus A at $T_{\min}$ shown in Fig. \ref{hysteresis} can be interpreted in terms of the changes in $r$.  At sufficiently low temperatures ($T<50$ mK), Eq. (\ref{r_atten_fit}) reduces to $\alpha_d = r \alpha_1$ because $\alpha_2=0$.  Similarly, Eq. (\ref{r_dv_fit}) reduces to $\delta v_d/v=r(\delta v_d/v)_1$.  As $A$ is decreased from $A_f$ of the previous cooling run, $r$ decreases according to Eq. (\ref{r-theory}) with $A_f$ replaced by $A$. 
When $A < A_c(\approx 1.54)$, $r$ becomes zero so that $\alpha_d = 0$ and $\delta v_d/v=0$.  As $A$ is now increased in the range $A< A_t(\approx 8)$, $r$ remains zero. When $A$ exceeds $A_t$, catastrophic unpinning of dislocations occurs and $r$ increases. It is this unpinning process that shows long relaxation. Thus the origin of the hysteretic behavior at $T_{\min}$ shown in Fig. \ref{hysteresis} can be traced to the hysteresis in $r$.

\section{Discussion}\label{Discussion}
\subsection{Relation and brief comparison with earlier ultrasound experiments on solid $^4$He}
The present ultrasound results in the higher temperature range ($T>200$ mK) are fairly well understood as a transition from the overdamped to underdamped resonance of dislocations by the Granato-L\"{u}cke theory.  These results are similar to the previous ultrasound measurements on single crystalline samples of $^4$He\cite{Iwasa79,Wanner76}, although our samples are likely polycrystalline.  The similarity comes about owing probably to the size of crystal grains\cite{Schuch62,Wilks67} (typically 1 mm) in our samples being much larger than the pinning length of dislocation segments (typically 10 $\mu$m).

Our observations show that the dislocations are pinned by $^3$He impurities in the lower temperature range ($T<200$ mK).  Such observations were not reported in the previous ultrasound studies on single-crystalline solid $^4$He\cite{Iwasa79,Wanner76} grown from commercial $^4$He gas .  Probably, the temperature was not sufficiently lowered or the drive level was too high in the early measurements. The pinning effect is most likely present also in single-crystalline solid $^4$He grown from commercial $^4$He gas when the temperature is lowered down to 15 mK with sufficiently low drive level.

In one of the early measurements\cite{Lengua90}, the temperature was lowered below 100 mK but no dislocation effects, especially no pinning effects were reported because either the dislocation contributions to the ultrasound speed and attenuation were too small or the $^3$He concentration was too low (the authors used a higher purity gas with $x_3$= 5 ppb).  They observed contributions to the speed and attenuation from thermally activated elementary excitations and an additional resonant attenuation.  Our observations did not show such effects.

Ho, Bindloss and Goodkind\cite{Ho97} reported ultrasound measurements on solid $^4$He containing 27.5 ppm of $^3$He.  They observed a new anomaly at about $T_p$ = 165 mK characterized by a sharp attenuation peak and an increase in the speed of sound with decreasing temperature.  They analyzed the attenuation peak and the accompanied speed change as a relaxation mechanism phenomenon, and explained the anomaly as due to a continuous phase transition (second order phase transition), suggesting supersolidity\cite{Ho97}. In our measurements on samples with $x_3$ = 20 ppm (see Fig. \ref{H0e-atten-dv-38db}), a somewhat similar increase in speed of sound was observed at temperatures below 150 mK, but the attenuation did not show a peak in contrast to their data.  Our observations below 150 mK can be described in terms of pinning of dislocations by $^3$He impurities just like in our other samples with lower $^3$He concentration.  The characteristic temperatures $T_1$ and $T_2$ would be shifted higher in the $x_3$ = 20 ppm sample since the average impurity pinning length at a given temperature becomes shorter as $x_3$ is increased. If the anomaly observed by Ho et al. is related to the dislocation pinning, amplitude dependence should be observed on the lower side of $T_p$.  They, on the contrary, observed amplitude dependence on the higher side between $T_p$ and 500 mK. Hence, the anomaly does not seem to be caused by the dislocation pinning.

Iwasa and Suzuki \cite{Iwasa80} reported ultrasound measurements on single-crystalline $^4$He containing various concentrations of $^3$He; $x_3$=30 ppm, 300 ppm, and 1 \%.  They observed nonlinear (amplitude-dependent) attenuation and speed of sound between 120 and 900 mK in the sample with $x_3$=30 ppm.  They argued that the nonlinearity was related to pinning and unpinning of dislocations by $^3$He impurities.  This is consistent with our present results.  The higher onset temperature of nonlinearity (900 mK) in their measuerements at $x_3$=30 ppm compared with the onset temperature (150 mK) in our $x_3$=20 ppm sample possibly indicates that the actual $^3$He concentration in one or both of the samples may be substantially different from the given values.

\subsection{Comparison with shear modulus measurements}\label{ShearModulus}
It is interesting to compare our ultrasound results with those of shear modulus.  Haziot et al.\cite{Haziot13b} find that their shear modulus measurements and analysis of polycrystalline $^4$He samples show a dislocation density of $5.4\times 10^9$ m$^{-2}$ with an average length of 59 $\mu$m.   Our $R\Lambda$ is consistent with their dislocation density, considering that the orientation factor of a randomly oriented polycrystalline sample is 1/4 for the shear wave and 1/16 for the longitudinal ultrasound.  Our  $L_{nA}$, on the other hand, is smaller by one order of magnitude than their length. This discrepancy may arise from the number of pinning points affecting the ultrasound propagation being much greater than in the shear modulus experiment. Since the network nodes act as pinning points in both shear modulus and ultrasound experiments, there must be additional pinning points in the ultrasound experiment.  A candidate for the additional pinning points is jogs, but we do not have sufficient information on the microscopic structure of basal dislocations and jogs for further analysis.

Fefferman et al. \cite{Fefferman14} reported the binding energy of $^3$He atoms on dislocations as $E_b=0.67$ K and the critical force $F_c = 6.8\times 10^{-15}$ N.  Their binding energy is about a factor of two greater than ours.  The discrepancy may be due to the different models in the analysis.  
The binding energy in the present work is obtained from the temperature dependences of the sound speed and attenuation assuming a T-dependent impurity pinning length (pinning model).
The binding energy in the shear modulus measurements, on the other hand, was obtained from the frequency dependence of the dissipation-peak temperature assuming that the damping force on dislocation motion at low temperature was proportional to the concentration of $^3$He bound to the dislocations (damping model).  
If $E_b=0.67$ K is assumed in the analysis of the attenuation and speed of sample $\#11$ at $A_f=1.12$, the results of fitting are shown as red broken curves in Fig. \ref{binding_energy}.  As can be seen, the variation of the curves between 60 and 300 mK differs from the measurement because $L_{iA}$ with $E_b=0.67$ K varies faster than that with $E_b=0.35$ K.

Since $E_b$ is the depth of the spatially dependent binding potential and $F_c$ is its maximum slope, $E_b$ and $F_c$ are simply related to each other as shown in Appendix \ref{AppendixB}. Entering Fefferman et al.'s value of $F_c$ into Eq. (B4) gives $E_b = 0.23$ K.  This binding energy is different from their own estimate but close to ours.

Kang et al.\cite{Kang13} reported observations on the hysteresis of shear modulus while scanning the temperature and the applied stress.  Their measurement procedure did not allow observations
of thermal hysteresis like in ours.  On the other hand, they did observe stress-dependent hysteresis and explained it in terms of the pinning/unpinning of dislocations by $^3$He impurities similar to our analysis.  They assumed distributions in the impurity pinning length as well as in the network pinning length.\cite{Kang15} 

Although dislocation effects can be observed in both shear modulus and ultrasound measurements, significant differences should be noted. The ultrasound is selectively sensitive to those dislocation segments with lengths close to $L_0$.  This is advantageous since $L_{nA}$ tends to be close in range to $L_0$ in many samples.  The shear modulus is sensitive to those with much broader range of lengths. Spatial variation in the stress amplitude plays an important role in ultrasound, while the stress amplitude in the shear modulus experiment can be assumed to be uniform owing to the much longer wavelength.

\subsection{Small-attenuation response}\label{sa-response}
There are several possible effects that would produce the small-attenuation response shown in Fig. \ref{D1b-atten-dv}; (1) low dislocation density, (2) small orientation factor, (3) short network pinning length due to high dislocation density, (4) suppression of dislocation motion due to high impurity concentration, etc.  A small orientation factor is possible for a single crystal when the angle between the c-axis and the sound propagation direction, $\theta$, is equal to 0 or 90$^{\circ}$ (the  orientation factor of longitudinal ultrasound in a single-crystal hcp $^4$He is given\cite{Iwasa79} by $R=(1/8)\sin ^2 2\theta$ ). The small-attenuation response of a single crystal is reported in Fig. 3 in Calder and Franck\cite{Calder77} which may be due to the effect (2) above.  The small-attenuation response due to the effect (4) is observed for a single crystal doped with 1 $\%$ $^3$He.\cite{Iwasa80} However, the real origin of the small-attenuation response of the present likely polycrystalline samples is not clear.

\subsection{Anomalous response}\label{anomalous discussion}
The sample showing anomalous response (see Fig. \ref{F1-attenuation-dv}) accompanied by an unexpected peak in attenuation and rapid change in speed near 700 mK is now discussed.  The response is suggestive of a Debye relaxation process.  The change in speed and attenuation according to the Debye relaxation model can be written as
\begin{equation}
\frac{\delta v_r}{v} =  -\frac{\phi}{1+(\Omega\tau_r)^2},
\label{v_r}
\end{equation}
and
\begin{equation}
\frac{\alpha_r}{x_m}=  -\frac{\phi}{v}\frac{\Omega^2\tau_r}{1+(\Omega\tau_r)^2},
\label{alpha_r}
\end{equation}
where $\phi$ is a constant and $\tau_r$ is a characteristic relaxation time.
Eq. (\ref{v_r}) is fitted to the $\delta v/v$ data of the sample $\#6$ around 700 mK as shown by the solid line in Fig. \ref{F1-attenuation-dv}(2) and the fitted relaxation time is $\tau_r = 2.2\times 10^{-4} \exp(-6.91/T)$.  The activation energy in $\tau_r$ is larger than $E_b$ and smaller than the formation energy of a vacancy (more than 10 K).  The position and height of $\alpha_r$ calculated from Eq. (\ref{alpha_r}) are in good agreement with the data as shown in Fig. \ref{F1-attenuation-dv}(1).  Although the anomaly is well described with the Debye relaxation model, its physical origin is not clear yet.

\subsection{Dependence on $^3$He impurity concentration}
Consider the variation in the onset temperatures $T_1$ and $T_2$ for the thermally-assisted stress-induced unpinning as $x_3$ is increased from 0.3 to 20 ppm.  Qualitatively, the average impurity length $L_{iA}$ decreases with increasing $x_3$ so that unpinning would become more difficult, and $T_1$ and $T_2$ would become higher.  Indeed $T_1$ does shift up from 50 mK to 80 $\sim$ 90 mK and $T_2$ from 70 mK to 100 $\sim$ 110 mK.  Quantitatively, it is expected that the onset occurs at the temperatures where the average impurity length $L_{iA}$ is the same in the two impurity concentrations.  In the $x_3 = 0.3$ ppm samples, Eq. (\ref{impurity_length2}) with $E_b$ = 0.35 K gives $L_{iA} = 0.11$ $\mu$m at the observed $T_1$ = 50 mK.  In the case of $x_3 = 20$ ppm, the same $L_{iA}$ occurs at 125 mK, which is fairly close to the observed value of $T_1$ for $x_3 = 20$ ppm. Similar calculation for $T_2$ gives 437 mK for $x_3 = 20$ ppm, which is much higher than the observed value. 
Note there are uncertainties in this estimation. (1) $L_{iA}$ may\cite{Iwasa80} be proportional to $x_3^{-2/3}$ instead of Eq. (\ref{impurity_length2}).  In this case $T_1$ and $T_2$ at $x_3=20$ ppm are expected to be 83 mK and 159 mK, respectively, in better agreement with the observation.  (2) The \textit{in-situ} value of $x_3$ may be different from 20 ppm.
More study is required for quantitative analysis.

\subsection{Relaxation phenomena}\label{relaxation_phenomena}
The relaxation phenomena as shown in Fig. \ref{hysteresis} and \ref{amp-dv-relaxation} are, as stated earlier, likely related to the dynamics of ``capture/release" of $^3$He impurities during pinning/unpinning of dislocations.  The processes of decreasing and increasing of $A$ involve distinct mechanisms and are discussed in order below.

The process of decreasing $A$ in Fig. \ref{hysteresis} leads to a decrease in $r$ according to Eq. (\ref{r-theory}) in which $A_f$ is replaced by $A$. As $A$ is decreased, pinning of dislocations by impurities occurs in the network-pinned region near $x = x_0$, where the stress amplitude due to ultrasound is the smallest. The response of ultrasound signal to the stepwise decreases in the drive amplitude is completed within less than a few seconds at $T_{\min}$ (see Fig. \ref{amp-dv-relaxation}(1)).  This indicates that impurities are captured by dislocations on this short time scale ($\sim$4 s).
It is also desirable to estimate the lower limit of the relaxation time for pinning.  Obviously, the relaxation time is longer than the repetition time of the ultrasound pulses (1 ms).  Otherwise the dislocations are pinned between the successive pulses and no changes in $\alpha$ and $\delta v/v$ would occur when $A$ is decreased.
According to Iwasa \cite{Iwasa13}, the impurity pinning rate $R_1$ is independent of $T$ and given by 
\begin{equation}
R_1=L x_3 Q
\label{R1}
\end{equation}
where $L$ is the length of dislocation segment and $Q=1.0\times 10^{12}$ (m$\cdot$s)$^{-1}$ is a constant. For a resonant dislocation segment, $L=L_0 =7.5$ $\mu$m, and with $x_3=0.3$ ppm, the pinning rate is $R_1 = 3$ s$^{-1}$ and the relaxation time, $1/R_1 = 0.3$ s. This is consistent with the observed response time.
Corboz et al.\cite{Corboz08a}, on the other hand, theoretically considered the interaction between $^3$He impurity and screw dislocation and found that the impurity capture relaxation time would be in the order of hours and days.  Such a long relaxation time is not observed.

The process of increasing $A$ in Fig. \ref{hysteresis} is very different. Initially, the drive level is set to the small value $A_i$ and the dislocations in the entire sample are pinned to the maximum extent possible by impurities. When $A$ is increased to a value less than $A_t$, the dislocations remain pinned and the response is linear, i.e. $\alpha$ and $\delta v/v$ do not change and the signal $S$ increases in proportion to $A$ with a short response time (see Fig. \ref{amp-dv-relaxation}(2)). When $A$ is increased, however, beyond $A_t$, unpinning is initiated and the catastrophic unpinning follows.  The long relaxation time in Fig. \ref{amp-dv-relaxation}(3) indicates the unpinning as a stochastic and dynamical process. 

Immediately after the increase in $A$ but before any significant unpinning has occurred, the stress amplitude is uniform throughout the sample since the attenuation coefficient in the purely pinned state is negligibly small.  Unpinning can then be induced by ultrasound stress anywhere in the sample.  As the number of unpinned segments increases, the stress amplitude decreases along $x$ and the probability of unpinning also decreases.

The stochastic and dynamical unpinning process may be pictured as follows. The critical length for  $A=14.1$ is estimated to be $L_c=0.85$ $\mu$m from Eq. (\ref{Lc-numerical}).  According to Eq. (\ref{impurity_length3}), the average impurity pinning length becomes $L_{iA}=\gamma b$ at low temperature. If $\gamma=1$, all the lattice sites along the dislocation line are occupied by $^3$He atoms and unpinning is impossible. We believe that a repulsive force between $^3$He atoms results in $\gamma$ much bigger than unity.
Nevertheless, $L_{iA}$ at 15 mK is smaller than $L_c$ even with $\gamma=100$ as shown in Fig. \ref{impurity-length}.
According to the distribution of the impurity pinning lengths, Eq. (\ref{network-impurity-distribution}), some of the impurity segments may be much longer than $L_{iA}$.  In addition, individual impurity pinning length may fluctuate because $^3$He impurity atoms can move along the dislocation line at low temperetures due to quantum tunneling.  The condition of unpinning, Eq. (\ref{Lc-equation-2}), can be occasionally satisfied due to the fluctuation and the dislocation segment can be unpinned.
All these processes are likely involved in the observed slow relaxation shown in Fig. \ref{amp-dv-relaxation}(3). The relaxation time decreases to less than 10 s at 65 mK, but no systematic study on the temperature dependence of the relaxation time was made.

More work at temperatures between 10 and 100 mK is clearly needed in elucidating the relaxation phenomena in the unpinning process.
Possible experiments include (1) increasing $A_f$ systematically at $T_{min}$ to see the change in the unpinning time, (2) applying a sudden DC stress and observing the dislocation unpinning, and (3) a kind of pump-probe method in which a large amplitude RF pulse is applied on the transmitter and the subsequent ultrasound response is measured.

\subsection{Thermal effects at low temperatures}\label{v-change}
At temperatures below 200 mK in the underdamped regime, the damping of dislocation vibration due to the fluttering mechanism can be neglected.  The reversible temperature dependences of $\alpha$ and $\delta v/v$ at low drive level such as the data at $A=1.12$ in Fig. \ref{L1-2D-T-3drive-atten-dv-wc-v2} are due to pinning and unpinning of dislocations by $^3$He impurities which occurs uniformly in the whole sample irrespective of the magnitude of $L_N$.
These effects of temperature are quite different from those of stress that causes unpinning preferentially in the region at higher amplitude and for the dislocation segments with longer pinning length.

Thus the temperature dependences of $\alpha$ and $\delta v/v$ at low drive level originate from that of $L_{A}(T)$.  Numerical calculations of $\alpha_d$ and $L_{A}(T)$ using the parameters for the sample $\#11$ provide $\alpha_d = 0.1$ dB and $L_{A}(T) = 0.7$ $\mu$m at 74 mK; $\alpha_d = 1$ dB and $L_{A}(T) = 0.95$ $\mu$m at 80 mK; and $\alpha_d = 10$ dB and $L_{A}(T) = 1.7$ $\mu$m at 105 mK.
Similar calculation of $\delta v_d/v$ shows that a minimum of $\delta v_d/v = -1.1 \times 10^{3}$ occurs at 98 mK where $L_{A}(T) = 1.5$ $\mu$m.  It indicates that the velocity minimum occurs at $L_{A}/L_0=0.2$ when the exponential distribution is assumed.

Numerical value of $\delta v_d/v$ in the network-pinned state calculated with Eq. (\ref{v_d})
and (\ref{network-distribution}) shows that $\delta v_d/v < 0$ for $L_{nA} < 2.6$ $\mu$m corresponding to the case of the sample $\#$8 at temparatures above 150 mK (see Fig. \ref{H0e-atten-dv-38db}), while $\delta v_d/v > 0$ for $L_{nA} > 2.6$ $\mu$m corresponding to the case of the sample $\#$6 at temparatures above 100 mK (see Fig. \ref{F1-attenuation-dv}).

\section{Conclusion}\label{Conclusion}

A systematic study of 9.6 MHz ultrasound propagation in solid $^4$He was made between 1.2 K and 15 mK.  The attenuation and the changes in the speed of propagation were measured as functions of temperature and the drive amplitude.  Depending on the drive amplitude, the propagation characteristics showed linear, nonlinear, reversible and hysteretic behaviors.  Most of the observed behaviors in the “typical” samples could be explained in terms of the interaction between the ultrasound and the string-like vibration of dislocation lines that are strongly pinned at network nodes and weakly pinned by $^3$He impurity.  The small impurity concentration at the level of 0.3 ppm in the typical samples plays a crucial role in the ultrasound propagation at low temperature below 200 mK.

There are two extreme states; the network-pinned state at sufficiently high $A_f$ and the impurity-pinned state at sufficiently low $A_f$.  The temperature dependences of $\alpha$ and $\delta v/v$ in the network-pinned state can be described by the GL theory with the $T$-independent distribution of the dislocation pinning lengths similar to those in the single-crystalline $^4$He.  Those in the impurity-pinned state are described by the GL theory including the $T$-dependent pinning effects of dislocations by $^3$He impurity atoms.

In our simplified model, the sample is divided in two regions; the network-pinned region on the transmitter side and the impurity-pinned region on the receiver side, and the state of the sample is described with a single parameter $r$ which gives the $T$-, $A_f$- and history-dependent fraction of network-pinned region of the sample.

When the sample is cooled from 1.2 K, the value of $r$ depends on the drive amplitude $A_f$ but not on $T$.  When the drive level is decreased from $A_f$ to $A_i$ at $T_{\min}$, $r$ becomes 0.  When $A$ is increased at $T_{\min}$, $r$ remains 0 up to $A_f=A_t (\approx 8)$.  When $A_f > A_t$, $r$ starts to increase with a long relaxation time.

In warming runs when $A_f < A_c (\approx 1.5)$, $r$ remains 0 throughout and the attenuation increases at $T > T_2$ where thermal unpinning becomes appreciable.  When the value of $A_f$ is intermediate between $A_c$ and $A_t$, for example $A_f = 4$, $r$ remains 0 up to $T_1$.  Then thermally-assisted stress-induced unpinning starts at $T > T_1$ and $r$ starts to increase at the same time.  At temperatures $T > T_2$ thermal unpinning becomes dominant so that $\alpha$ continues to increase but $r$ stays constant.

The present work has raised questions about pinning and unpinning processes of dislocations by $^3$He impurities.  The origins of the short relaxation time for pinning as well as the long relaxation time for unpinning are yet to be clarified.  Measurement on the temperature dependence of the relaxation time may be helpful.  In order to get a more comprehensive picture, studies in the frequency range between 100 kHz and 1 MHz would be desirable.
Ultrasound measurements on single crystalline $^4$He samples at temperatures below 100 mK are also desirable in order to study the pinning mechanism of dislocations by $^3$He impurities.  Possible effects of grain boundaries in polycrystalline samples can be eliminated in single crystals.
Finally, a systematic study of the dependence on $^3$He impurity concentration including ultra high purity samples down to 10 mK and lower would be interesting to explore the extent of the validity of our interpretations.

\begin{acknowledgments}
We are most grateful to John Goodkind for providing us with the ultrasound instrumentation system utilized in the experiments.  We thank Elizabeth Eibling and Parth Jariwala for help in designing of apparatus, Bettina Hein, Rebecca Cebulka, Michelle Goffreda and Chirag Soni for data analysis, and Michael Keiderling for data acquisition.  We are grateful to the support and encouragement from Bob Bartynski.
This research was supported by NSF DMR-1005325.
\end{acknowledgments}

\setcounter{figure}{0}
\renewcommand{\thefigure}{A\arabic{figure}}

\appendix \label{Appendix}

\section{Calibration of transducers}\label{AppendixA}
\begin{figure}
\includegraphics[width=3.4in]{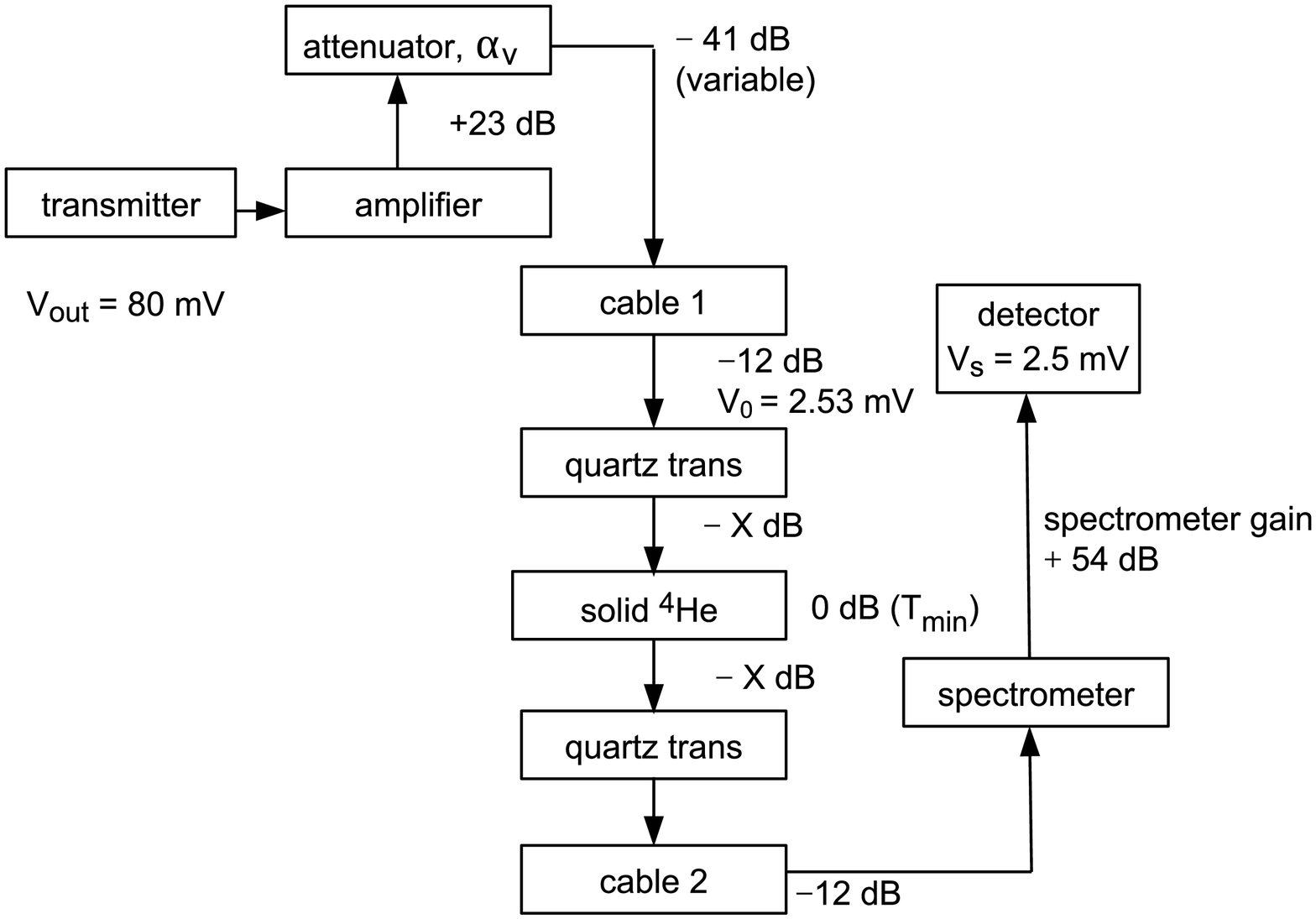}
\caption{
Energy diagram of the ultrasound system.  Losses and gains in the components in the measurement system are indicated.  The loss in the ultra-miniature coaxial cable was calculated to be 12 dB by measuring the input (80 mV peak-to-peak RF pulse) and output (20 mVp-p) through the cable.  $X$ is the transmission loss at each of the interfaces from drive quartz transducer to solid $^4$He and from solid $^4$He to receiver quartz transducer.  When the variable step attenuator is set to $\alpha_v = 41$ dB corresponding to $A = 1$, the applied voltage on the drive quartz transducer is $V_0 = 2.53$ mV.  The output signal amplitude of the spectrometer is $V_s = 2.5$ mV when the solid $^4$He sample is at our minimum temperature($\sim$15 mK).
}
\label{calibration}
\end{figure}

The relation between the ultrasound drive amplitude and the stress exerted onto the solid $^4$He samples is roughly calibrated by measuring the losses and gains in the measurement system components  as shown in Fig. \ref{calibration}.  The present estimation is, however, uncertain by as much as a factor of two due to incomplete impedance matching, acoustic mismatch, and ringing of the transducer in the ultrasound experiment.  When the amplitude of the transmitter output is $V_{out}=80$ mV and the variable step attenuator is arbitrarily set to $\alpha_v = 41$ dB, corresponding to the relative drive amplitude $A=1$, the voltage on the driver transducer is $V_0 = 2.53$ mV.
The spectrometer signal through the solid $^4$He sample cooled to near 15 mK is $V_s = 2.5$ mV.  Assuming the attenuation in the sample is negligible(see Sec. \ref{spatial variation}), we can relate the electrical input-output ratio to the gains and losses in the system in Fig. \ref{calibration} as $- 2X + 42\quad \mathrm{[dB]}= 20\log(V_s/V_0) =-0.1$ [dB], or $X = 21$ dB, where $X$ is the conversion loss at the transducer/solid $^4$He interfaces.

Consider a voltage pulse of amplitude $V$ and of time interval $\tau$.  The corresponding electrical pulse energy is $E_0 = (V^2/2 Z_0)\tau$ where $Z_0=50$ $\Omega$ is taken as the cable impedance.  Using $V/V_0=A$, we can write
\begin{equation}
E_0 = \frac{V_0^2}{2 Z_0}\tau A^2 .
\label{A-10}
\end{equation}
The mechanical pulsed energy $E_1$ transmitted from the quartz transducer into solid $^4$He is reduced by $X=21$ dB,
\begin{equation}
E_1=E_0 10^{-21/10}= 0.008 E_0 .
\label{A-11}
\end{equation}
The elastic energy is also written as
\begin{equation}
E_1 = \frac{C_l \epsilon_0^2}{2} \frac{\pi D^2}{4} v \tau ,
\label{A-1}
\end{equation}
where $C_l = 5.6\times 10^7$ Pa is the longitudinal modulus of hcp $^4$He, $\epsilon_0$ is the strain amplitude, $D = 8.6$ mm is the diameter of the sample, and $v = 534$ m/s is the longitudinal speed of sound.  From Eqs. (\ref{A-10}) through (\ref{A-1}) it is found that $\epsilon_0 = 2.4\times 10^{-8} A$ and the corresponding compressional stress amplitude is $\sigma_0 = C_l\epsilon_0 = 1.36 A$ Pa.   Since the sample is probably polycrystalline and each grain is randomly oriented, the relative angle between the sound propagation axis and the $c$-axis on average is arbitrarily taken as $\theta=45^{\circ}$.  Then, the shear stress amplitude($\sigma_s$) becomes, 
\begin{equation}
\sigma_s = \sigma_0\sin\theta\cos\theta = 0.68 A \quad \mathrm{Pa} ,
\label{A-2}
\end{equation}
and the corresponding shear strain amplitude($\epsilon_s$) is 
\begin{equation}
\epsilon_s = \frac{\sigma_s}{C_s}=4.8\times 10^{-8} A ,
\label{A-3}
\end{equation}
where $C_s = 1.4\times 10^7$ Pa is the shear modulus.  The shear stress amplitude calibration is used in estimating the critical force for unpinning of dislocation lines(see Appendix \ref{AppendixB}).

\section{Calculation of $r(A_f)$} \label{AppendixB}
We assume at first that the sample is a uniform medium with the attenuation coefficient $\beta$.  Ultrasound pulses are excited at $x=0$ and propagate in the $x$-direction.  The local amplitude of the pulses are given by
\begin{equation}
A(x)=A_f \exp(-\beta x) ,
\label{B-1}
\end{equation}
where $A_f$ is the relative drive amplitude at $x=0$.
The local shear stress $\sigma_s(x)$ is proportional to $A(x)$ (see Eq. (\ref{A-2})) and the local critical length $L_c(x)$ is inversely proportional to $\sigma_s(x)$ (see Eq. (\ref{Lc-equation})), so that 
$L_c(x)$ is inversely proportional to $A(x)$;
\begin{equation}
L_c(x)=\frac{q}{A(x)}=\frac{q}{A_f} \exp(\beta x) ,
\label{B-2}
\end{equation}
where
\begin{equation}
q=\frac{2F_c}{b \sigma_s(x)} A(x) = \frac{2F_c}{0.68b} .
\label{B-2b}
\end{equation}
Eq. (\ref{A-2}) is used in the second equality.

We next assume that the position, $x_0$, of the boundary between network- and impurity-pinned regions is determined by the condition
\begin{equation}
L_c(x_0)=L_0 .
\label{B-3}
\end{equation}
From Eq. (\ref{B-2}) and (\ref{B-3}), $x_0$ and $r$ are determined to be
\begin{equation}
x_0=\frac{1}{\beta}\ln\frac{A_f L_0}{q} 
\label{B-4}
\end{equation}
and
\begin{equation}
r(A_f)=\frac{x_0}{x_m}=\frac{1}{\beta x_m}\ln\frac{A_f L_0}{q} .
\label{B-5}
\end{equation}
Note the range of $r$ is between 0 and 1, so that $r=0$ for $A_f \leq q/L_0$ and $r=1$ for $A_f \geq (q/L_0) \exp(\beta x_m)$.

When we fit the data of $r$ except for $A_f=1.12$ shown in Fig. \ref{r-Af} with Eq. (\ref{B-5}), we obtain
\begin{equation}
r(A_f)=\frac{1}{2.5}\ln\frac{A_f}{1.54} .
\label{B-6}
\end{equation}
Comparing Eq. (\ref{B-6}) with Eq. (\ref{B-5}), we obtain $q=1.54 \cdot L_0=1.2\times 10^{-5}$ m,
\begin{equation}
L_c(x) [\mbox{m}] = \frac{1.2\times 10^{-5}}{A(x)} 
\label{B-9}
\end{equation}
from Eq. (\ref{B-2}), and $F_c=1.5\times 10^{-15}$ N from Eq. (\ref{B-2b}).  We also obtain $\beta x_m=2.5$ which leads the attenuation in the uniform sample with the attenuation coefficient $\beta$ and length $x_m$ to be
\begin{equation}
20 \log \frac{A_f}{A(x_m)}=21.7 \mbox{dB} .
\label{B-7}
\end{equation}
This value is consistent with the attenuation in the network-pinned state shown in Fig. \ref{atten-dv-pinned-unpinned-limits}, $\alpha_1(T_{\min}) = 23.2$ dB.

The numerical values in Eq. (\ref{B-9}) and (\ref{B-7}) are rather reliable.  The numerical value of $F_c$, on the other hand, is likely accurate only to an order of magnitude owing to various uncertainties in deriving Eq. (\ref{A-2}).

\section{Relation between $E_b$ and $F_c$} \label{AppendixC}

The interaction energy between an edge dislocation lying along the $z$-axis at (0,0) and a $^3$He impurity atom at $(x,y)$ is given by\cite{Hull65}
\begin{equation}
W(x,y)=\frac{4(1+\nu)}{3(1-\nu)}\mu b_e {r_a}^3 \delta \frac{y}{x^2+y^2} ,
\label{C-1}
\end{equation}
where $x$ and $y$ are the coordinates of the impurity parallel and perpendicular to the slip plane, respectively, $\nu$ is the Poisson's ratio, $\mu$ is the shear modulus, $b_e$ is the edge component of the Burgers vector, $r_a$ is the atomic radius of a $^4$He atom, $r_a(1+\delta)$ is that of a $^3$He atom, and $\delta$ is the misfit parameter.  We note $b_e=b$ for a perfect edge dislocation on the basal plane and $r_a=b/2$, where $b$ is the lattice parameter.  The most stable position of the impurity is $(0,-b)$ so that the binding energy is given by
\begin{equation}
E_b=W(\infty,-b)-W(0,-b)=\frac{(1+\nu)}{6(1-\nu)}\mu b^3 \delta .
\label{C-2}
\end{equation}
The ciritical force is given by 
\begin{equation}
F_c=\left( \frac{\partial}{\partial x} W(x,-b) \right)_{\max}=\frac{\sqrt{3} (1+\nu)}{16(1-\nu)}\mu b^2 \delta ,
\label{C-3}
\end{equation}
which occurs at $x=b/\sqrt{3}$.
We obtain from Eqs. (\ref{C-2}) and (\ref{C-3}),
\begin{equation}
F_c=\frac{3\sqrt{3}}{8}\frac{E_b}{b} \approx 0.65\frac{E_b}{b}.
\label{C-4}
\end{equation}
Fefferman et al. \cite{Fefferman14} mention $F_c \approx E/4b$ in contrast with Eq. (\ref{C-4}).

\bibliographystyle{apsrev4-1}
\bibliography{supersolid_v3,material,dislocation,helium4,technique}

\end{document}